\documentclass{aa}
\usepackage{txfonts}
\usepackage{natbib}
\bibpunct{(}{)}{;}{a}{}{,}
\usepackage{graphicx}
\usepackage{verbatim}
\usepackage[normalem]{ulem}
\usepackage{url}

\title{On the relation between metallicity and RGB color in HST/ACS data}
\author{D. Streich\inst{\ref{AIP}}
\and R. S. de Jong\inst{\ref{AIP}}
\and J. Bailin\inst{\ref{UA}}
\and P. Goudfrooij\inst{\ref{STScI}}
\and D. Radburn-Smith\inst{\ref{UW}}
\and M. Vlajic\inst{\ref{AIP}}
}
\institute{Leibniz-Institut f\"ur Astrophysik Potsdam (AIP), An der Sternwarte 16, 14482 Potsdam, Germany, \email{dstreich@aip.de}\label{AIP}
\and Department of Physics \& Astronomy, University of Alabama, Box 870324, Tuscaloosa, AL 35487, USA\label{UA}
\and Space Telescope Science Institute, 3700 San Martin Drive, Baltimore, MD 21218, USA\label{STScI}
\and Department of Astronomy, University of Washington, Box 351580, Seattle, WA 98195, USA\label{UW}
}
\date{Received .../ Accepted ...}  

\abstract{The determination of stellar metallicity and its gradient in external galaxies is a difficult task, but crucial for the understanding of galaxy formation and evolution.}
{The color of the Red Giant Branch (RGB) can be used to determine metallicities of stellar populations that have only shallow photometry. We will quantify the relation between metallicity and color in the widely used HST ACS filters F606W and F814W.}
{We use a sample of globular clusters from the ACS Globular Cluster Survey and measure their RGB color at given absolute magnitudes to derive the color-metallicity relation. We especially investigate the scatter and the uncertainties in this relation and show its limitations.}
{There is a clear relation between metallicity and RGB color. 
A comparison with isochrones shows reasonably good agreement with BaSTI models, a small offset to Dartmouth models, and a larger offset to Padua models.}
{Even for the best globular cluster data available, the metallicity of a simple stellar population can be determined from the RGB alone only with an accuracy of 0.3\,dex for [M/H]$\lesssim-1$, and 0.15\,dex for [M/H]$\gtrsim-1$. For mixed populations, as they are observed in external galaxies, the uncertainties will be even larger due to uncertainties in extinction, age, etc. Therefore caution is necessary when interpreting photometric metallicities.} %

\begin{document}
\maketitle

\section{Introduction}

Measuring the metallicity and its gradients in galaxies is a key issue for understanding galaxy formation and evolution. 

Outside the Local Group, spectroscopic metallicity determination of (resolved) stars is not feasible at the moment, except for the few very bright supergiants \citep{kudritzki12}. At the same time, the number of available color magnitude diagrams (CMDs) of nearby galaxies is increasing rapidly, e.g. from the GHOSTS \citep{radburn11} and ANGST \citep{dalcanton09} surveys.
These CMDs can be used to derive metallicities.

The color of the red giant branch (RGB) has long been known to depend on the metallicity \citep{hoyle55,sandage66,demarque82}. This has been extensively used to measure metallicities of old populations. 

There are many ways to convert the color to a metallicity: some authors define indices of an observed population, e.g. the color of the RGB at a given magnitude or the slope of the RGB, \citep[as defined for example in][]{dacosta90,lee93,saviane00,valenti04}, while others measured metallicities on a star by star basis by interpolating between either globular cluster fiducial lines \citep[e.g.][]{tanaka10,tiede04} or analytic RGB functions \citep[calibrated with globular cluster data, e.g. ][]{zoccali03,gullieuszik07,held10} or stellar evolution models \citep[e.g.][]{richardson09,babusiaux05}, and therefore generating metallicity distribution functions for a population. 

Uncertainties that arise from a specific calibration or a given isochrone or cluster template set are typically not well studied. Furthermore, an observational relation for the widely used HST ACS filters F606W-F814W is still missing in the literature\footnote{\citet{mormany05} actually have found such a relation, but they used only three clusters and did not publish the details.}. Here we aim to address these shortcomings.

This paper is organized as follows: After an introduction to the data and isochrones we use in chapter~\ref{data}, an observational color metallicity relation is derived in chapter~\ref{results}.
A discussion and summary follow in chapters~\ref{discussion} and~\ref{summary}.

\section{Data and Isochrones}
\label{data}

In this work we use the data of 71 globular clusters observed as part of the ACS Globular Cluster Survey \citep[ACSGCS;][]{sarajedini07} and its extension \citep{dotter11}. These data contain photometry in the F606W and F814W filters and is publicly available at the homepage of the ACSGCS team\footnote{\url{http://www.astro.ufl.edu/~ata/public_hstgc/}}. For the determination  of photometric uncertainties and completeness, the results from artificial star tests are also available. A detailed description of the data reduction is given in \citet{anderson08}.

To compare the different clusters, it is necessary to transform the apparent magnitudes into absolute, reddening-free magnitudes. For this purpose, we use the distance modulus and color excess from the GC database of W.~Harris \citep{harris96,harris10} and the extinction ratios for the ACS filters given by \citet[][Table 14]{sirianni05}. Metallicities, metallicity uncertainties and $\alpha$-abundances are taken from \citet{carretta09,carretta10}, if not stated otherwise.

In order to measure the color of the clusters RGBs they must have a sufficient number of stars in the RGB region. We selected therefore only those clusters for our study, which have more than five stars brighter than $M_{F814W}=-2$ and least one star brighter than $M_{F814W}=-3$. A list of the clusters used is given in Appendix~\ref{app:properties} (Table~\ref{GC_literature}).

For comparison with theoretical models, we use four sets of isochrones: the new PARSEC isochrones \citep{bressan12} and their predecessors, the (old) Padua isochrones\footnote{\url{http://stev.oapd.inaf.it/cgi-bin/cmd}} \citep[][and references therein]{girardi10,marigo08}, the BaSTI isochrones\footnote{\url{http://albione.oa-teramo.inaf.it/}} \citep{pietrinferni06,pietrinferni04} and the Dartmouth isochrones\footnote{\url{http://stellar.dartmouth.edu/~models/index.html}} \citep{dotter07}.

\section{Results}
\label{results}

\subsection{Color measurement}
\label{sec:colormeasurement}
We use two indices to define the color of the RGB: $C_{-3.0}=(F606W-F814W)_{M=-3.0}$ and $C_{-3.5}=(F606W-F814W)_{M=-3.5}$, i.e. the color of the RGB at an absolute F814W magnitude of -3.0 and -3.5, respectively (see Fig.~\ref{CMD_diverse} for some typical CMDs). Equivalent indices for the Johnson-Cousins filter system were already used by \citet{dacosta90}, \citet{lee93} and also by \citet{saviane00}. These indices have the advantage of only depending on relatively bright stars and can therefore be measured in distant galaxies, as well. We use also a third index, the S-index, which is the slope of the RGB \citep{saviane00,hartwick68}. This slope is measured between two points of the RGB, one at the level of the horizontal branch and the other two magnitudes brighter. While this index needs deeper data, and therefore its usage in extragalactic systems is limited, it has the advantage of being independent of extinction and distance errors.

In order to provide a robust measurement of the color at a given magnitude we interpolated the RGB with a hyperbola of the form:
$$ M=a+b\cdot \textrm{color}+c/(\textrm{color}+d)$$
Such a function was already used by \citet{saviane00} to find a one-parameter representation of the RGB; they defined the parameters $a$, $b$, $c$, and $d$ as a quadratic function of metallicity. Here, we are only interested in a good interpolation in sparse parts of the RGB and can therefore use $a$, $b$, $c$, and $d$ as free parameters for each cluster. In order to reduce problems due to contamination, we define a region of probable RGB stars, which also excludes the horizontal branch/red clump part of the CMD. Note that we fit the curve directly to the color/magnitude points of the stars and not to the ridge line of the RGB \citep[in contrast to][]{saviane00}. More details of the fitting process and some example plots with the exclusion region are shown in the Appendix.

To calculate the S-index, we first determined the horizontal branch magnitude of each system by visual inspection of the associated CMDs. This was typically F606W$\approx$0.40\,mag, with a 1-sigma variation of 0.10\,mag. We measured the color at this magnitude (and at 2 magnitudes brighter) from the fitted RGB used previously, and calculated the S-index as the slope between these points

\subsection{Metallicity determination}
The iron abundance [Fe/H] is often used synonymously with metallicity. However, from the theoretical point of view of stellar evolution, all elements are important in determining the properties of stellar atmospheres. Therefore the color of red giants is expected to depend on the overall metallicity [M/H] rather than on [Fe/H]. Unfortunately, there are very few measurements of the abundances of other elements in globular clusters.

\begin{figure}[!hb]
\centering
 \includegraphics[width=0.99\columnwidth]{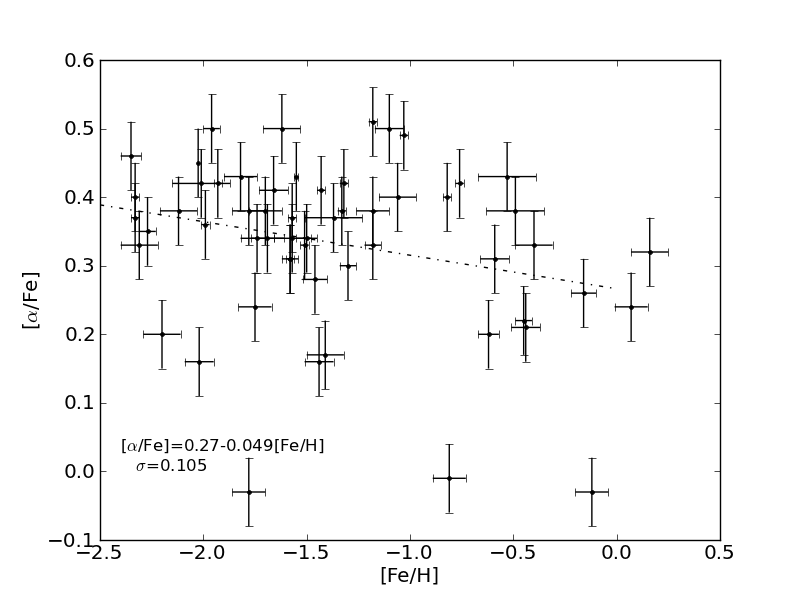}
 \caption{Alpha abundance as a function of [Fe/H] for all clusters in \citet{carretta10}. The text in the lower left corner gives the formula of the regression line and the scatter around this line. We used these for estimating the [$\alpha$/Fe] and its uncertainty for clusters without individual alpha measurement.}
\label{alphas}
\end{figure}

We use here the abundances given in \citet{carretta10}, who have measured [Fe/H] for all GCs in our sample and have compiled [$\alpha$/Fe] values for many of them. According to \citet{salaris93}, these two measurements can be combined to get the overall metallicity with the formula
$$[M/H] = [Fe/H] + \log_{10}(0.638*10^{[\alpha/Fe]}+0.362) .$$
For clusters that have no individual $\alpha$ measurement, we had to estimate its $\alpha$ abundance. Since the spread of [$\alpha$/Fe] among globular clusters is rather small, such an estimate will only introduce small errors. In Fig.~\ref{alphas}, [$\alpha$/Fe] is plotted against [Fe/H], where we have assumed an uncertainty of $0.05$ in the $\alpha$ abundance. The straight line is a linear regression, which we use for the estimation of [$\alpha$/Fe], where it is not available. The scatter around this regression line is 0.1\,dex, which we adopt as the individual uncertainty in the estimated [$\alpha$/Fe].

\subsection{Uncertainties}

To determine the uncertainties of our color measurements, we performed a bootstrap analysis. The uncertainty in the fit is derived by fitting the RGB of 500 samples that are drawn randomly from the original data. Each re-sample has the same number of stars as the original sample, but may contain some stars multiple times while others are absent.

We also incorporated in the bootstraps a shift due to the uncertainties in extinction and distance. According to \citet{harris10}, the uncertainty in extinction is of the order 10\% in E(B-V), but is at least 0.01\,mag, while the uncertainty in distance modulus is 0.1\,mag.

The uncertainty in distance is important because we measure the color at a given absolute magnitude. This is particularly significant for the metal-rich clusters where the color of the RGB is strongly dependent on magnitude, as opposed to metal-poor clusters where the RGB is nearly vertical on a CMD. The resulting uncertainty in $C_{-3.5}$ ranges from approximately 0.01\,mag at [M/H]=-2 to approximately 0.1\,mag at [M/H]=-0.2.

The uncertainties in the metallicity are the sum of the uncertainties in [Fe/H] \citet[as given by][]{carretta09}, and in [$\alpha$/Fe], which we adopt as 0.05\,dex for clusters with individual alpha-abundance measurements and 0.1\,dex for clusters with estimated values.

\subsection{Color metallicity relation}
Using the colors and metallicities described above, we can now look at the color-metallicity relations.

\begin{figure*}[!ht]
\centering
 \includegraphics[width=0.99\textwidth]{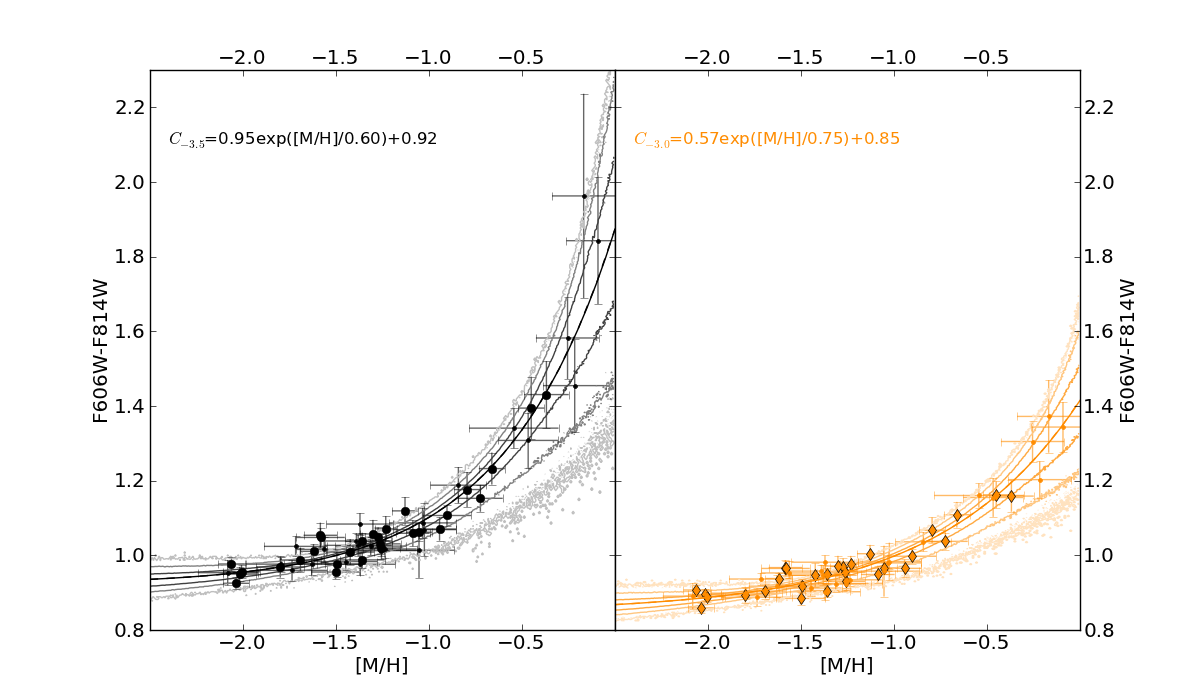}
 \caption{Color of the RGB as function of metallicity, based on the fitted RGBs. Black circles are for the color at $M_{F814W}=-3.5$, orange diamonds for $M_{F814W}=-3.0$. Small points are for clusters without individual alpha measurements. The solid lines are the best-fitting functions as given in the upper left corner and the lighter contours show the $1\sigma$, $2\sigma$ and $3\sigma$ regions of the fit.}
\label{metalfitindices}
\end{figure*}

The results are shown in Fig.~\ref{metalfitindices}. There is a clear relation between RGB color and spectroscopic metallicity. This relation can be parametrized with the function $F606W-F814W=a_0\exp(\mathrm{[M/H]}/a_1)+a_2.$ Using the orthogonal distance regression (ODR) algorithm \citep{boggs87,boggs92}\footnote{We used the Python implementation of this algorithm that is part of the Scipy library: \url{http://docs.scipy.org/doc/scipy/reference/odr.html}}, we determined the three parameters, that are shown in Table~\ref{fitparams}. The ODR uses the uncertainties on both variables to determine the best fit. Hence, both the uncertainties in color and metallicity, as described above, are considered during the fit and their effects are included in the final uncertainties of the resulting fit parameters.
The residual varinaces for both relations are $\sigma_{res}^2<1$, so the adopted uncertainties can explain the observed scatter in the relations.

\begin{table}[ht]
\caption{Fit parameters of the color--metallicity relations. For $C_{-3.5}$ and $C_{-3.0}$ the relation is exponential: $C_{i}=a_0\exp(\mathrm{[M/H]}/a_1)+a_2$, for the S-index it is linear $S=a_0+a_1\mathrm{[M/H]}$.
}
\label{fitparams}
\centering
\renewcommand{\arraystretch}{1.2}
\begin{tabular}{c|ccc}
	    & $a_0$ 	   & $a_1$ 		  & $a_2$ \\ \hline
$C_{-3.5}$&  $0.95\pm0.11$ & $0.602\pm0.069$ & $0.920\pm0.015$  \\
$C_{-3.0}$&  $0.567\pm0.056$ & $0.75\pm0.12$ & $0.845\pm0.018$ \\
S-index   &  $3.67\pm0.76$ & $-9.3\pm1.2$ &  $-2.08\pm0.44$  \\
\end{tabular}
\end{table}

\subsection{The S-index}
\begin{figure}[!ht]
\centering
 \includegraphics[width=0.99\columnwidth]{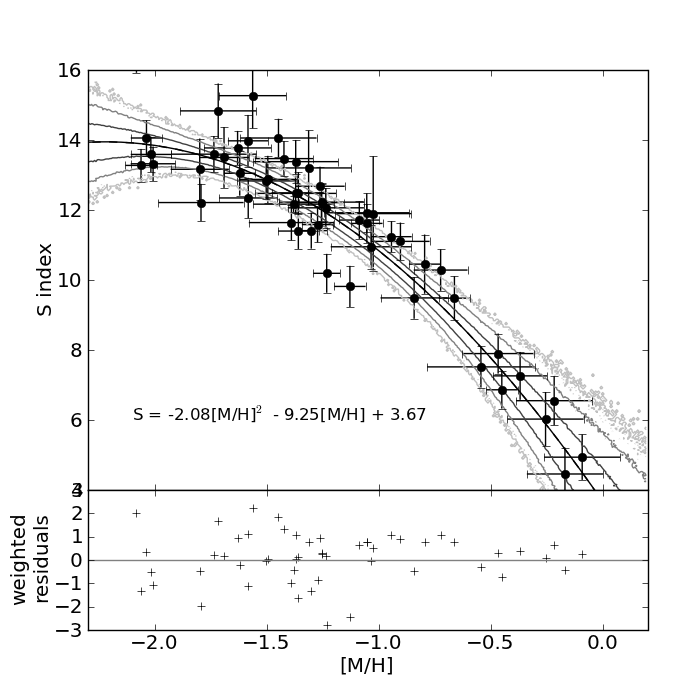}
 \caption{\textit{top panel:} The RGB slope as a function of metallicity. The solid black line is the best-fit quadratic function, as given in the equation in the bottom left. Grey lines give the $1\sigma$, $2\sigma$ and $3\sigma$ confidence ranges of the the best fit relation. \textit{Bottom panel:} weighted orthogonal residuals, i.e. the orthogonal distances to the best fit line divided by the respective uncertainties.
 }
\label{s_index}
\end{figure}

The slope of the RGB as a function of metallicity can be seen in Fig.~\ref{s_index}. 
The reported uncertainties of the S-index are a combination of the uncertainties of the RGB fit (determined through a bootstrap analysis as described above) and the uncertainty in the determination of the HB level, which we set here to $\sigma_{HBmag}=0.1$\,mag. 

As expected, the slope of the RGB gets smaller with increasing metallicity, while at the low-metallicity end the RGB slope is insensitive to metallicity. We have fitted a quadratic function to the data, which is shown in Fig.~\ref{s_index} together with the associated best-fit parameters. The choice of a quadratic function for the fit proves to be appropriate as no trends are seen in the residuals. Moreover, the variance of the residuals is only $\sigma_{res}^2=1.17$, i.e. the residuals are only slightly larger than expected from the individual measurement uncertainties. The maximum of the parabola is at [M/H]=-2.14, which is beyond the metallicity range of the observed clusters.

\section{Discussion}
\label{discussion}

\subsection{Analyzing residuals}

\begin{figure}[ht]
\centering
 \includegraphics[width=0.99\columnwidth]{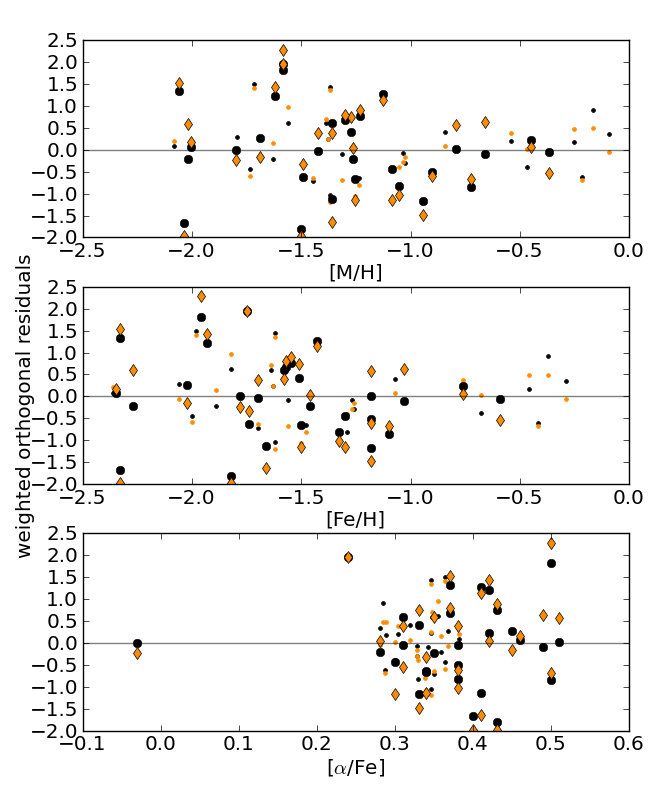}
 \caption{Residuals of the fit of the color-metallicity relation as function of metallicity (upper panel), iron abundance (middle panel) and alpha enhancement (lower panel). Symbols and colors are as in Fig.~\ref{metalfitindices}.}
\label{residuals1}
\end{figure}

\begin{figure}[ht]
\centering
 \includegraphics[width=0.99\columnwidth]{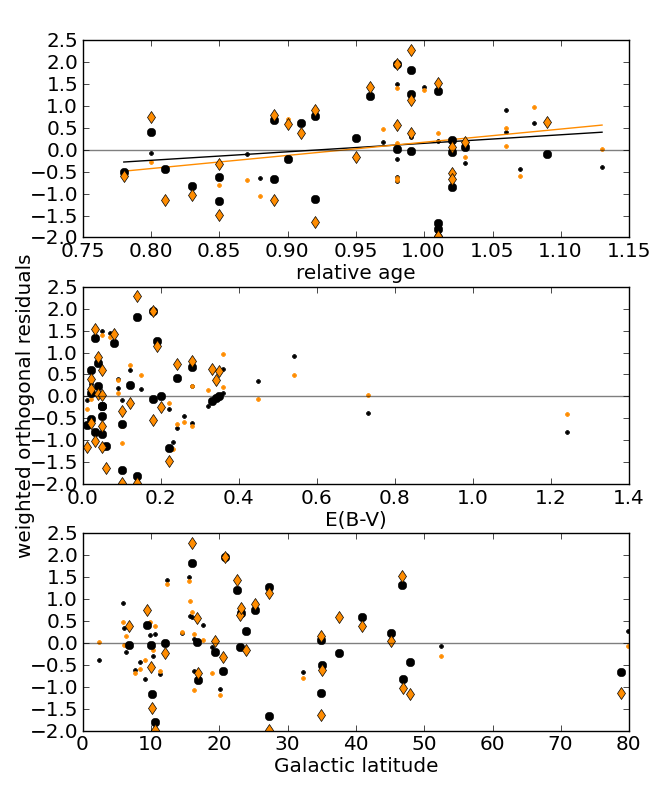}
 \caption{Residuals of the fit of the color-metallicity relation as function of age \citep[upper panel,][]{marinfranch09}, extinction (middle panel) and galactic latitude (lower panel). Symbols and colors are as in Fig.~\ref{metalfitindices}.}
\label{residuals2}
\end{figure}

\begin{figure}[ht]
\centering
 \includegraphics[width=0.99\columnwidth]{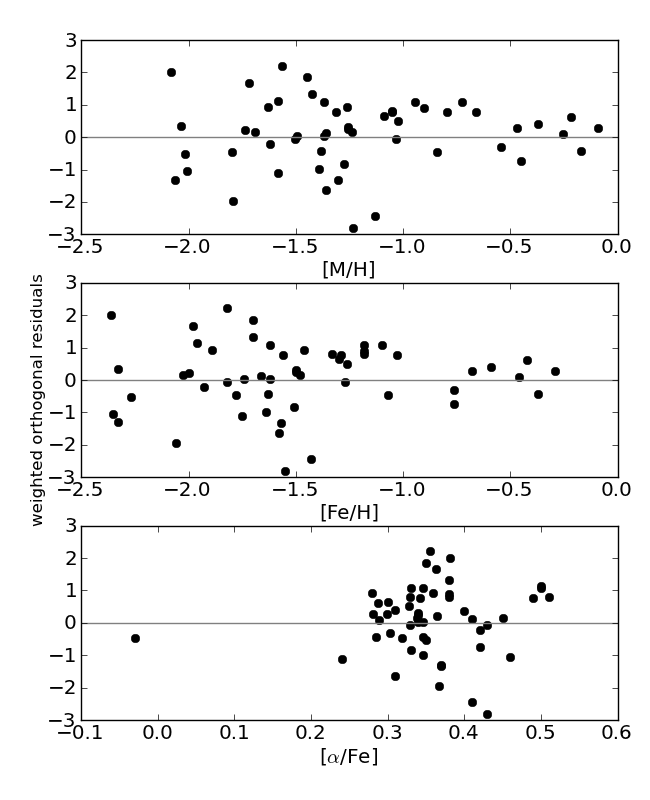}
 \caption{Residuals of the fit of the slope-metallicity relation as function of metallicity (upper panel), iron abundance (middle panel) and alpha enhancement (lower panel).}
\label{S_residuals1}
\end{figure}

\begin{figure}[ht]
\centering
 \includegraphics[width=0.99\columnwidth]{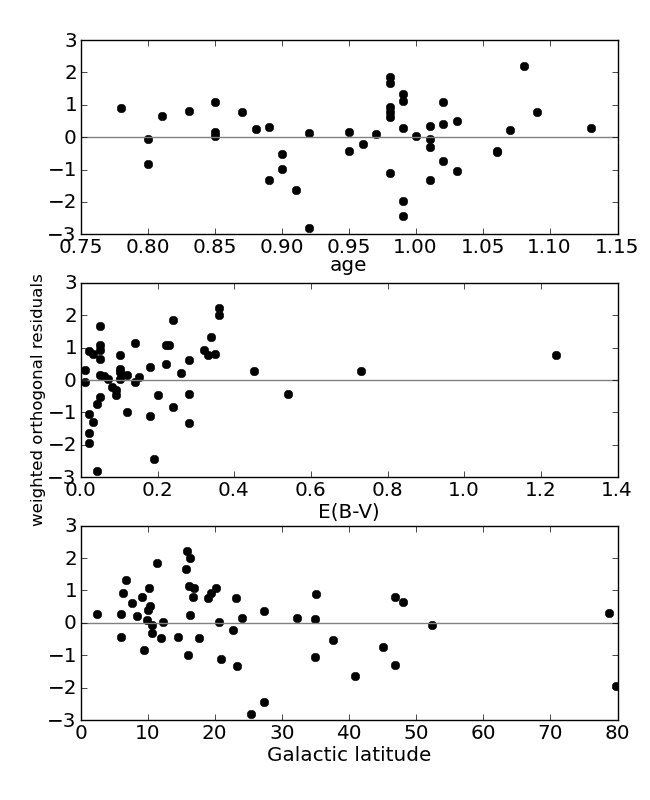}
 \caption{Residuals of the fit of the slope-metallicity relation as function of age \citep[upper panel,][]{marinfranch09}, extinction (middle panel) and galactic latitude (lower panel).}
\label{S_residuals2}
\end{figure}

We examine the residuals to look for a possible second parameter that influences the color or slope of the RGB and could produce some scatter in a simple color-metallicity relation. Figures~\ref{residuals1} and~\ref{residuals2} show the residuals of the fit of the color-metallicity relation, that is shown in Fig.~\ref{metalfitindices}, and Figures~\ref{S_residuals1} and~\ref{S_residuals2} the residuals of the fit to the slope metallicity relation, that is shown in Fig.~\ref{s_index}.

The residuals as a function of metallicity, [Fe/H] and [$\alpha$/Fe], are shown in Fig.~\ref{residuals1} and Fig.~\ref{S_residuals1}. There is no trend with any of these parameters, neither in the color- nor slope-metallicity relations. 

\begin{figure}[ht]
\centering
 \includegraphics[width=0.99\columnwidth]{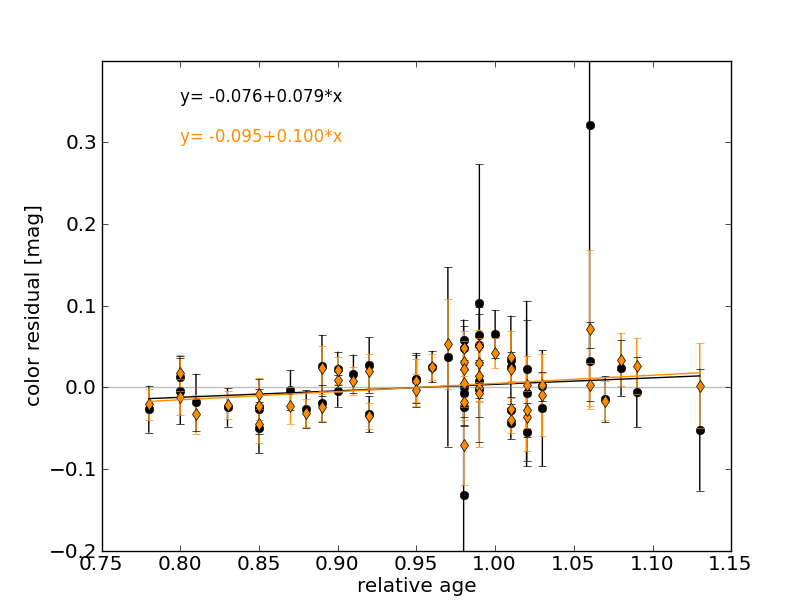}
 \caption{Color residuals of the fit of the color-metallicity relation as function of age. The text gives the regression line formulas for both indices.}
\label{color_residuals}
\end{figure}

From theoretical studies, the age is known to have an effect on the color of the RGB. 
In fact, a weak trend of the residuals with age can be seen in Figure~\ref{residuals2} (upper panel), with older clusters being slightly redder than younger clusters. The slope of the regression lines shown there is $1.94\pm2.13$ for $C_{-3.5}$ and $3.00\pm1.85$ for $C_{-3.0}$, which makes the trend significant for the $C_{-3.0}$ index. In order to quantify the effects of age on the CMD, we analyse the residuals in color space\footnote{i.e. we ignore uncertainties in metallicity and only look at the color offset between the data and the best fit relation} in Figure~\ref{color_residuals}.
Assuming a typical age for globular clusters of 12.8\,Gyr (as \citet{marinfranch09} do using the isochrones of \citet{dotter07}) we can transform the slopes of the regression lines in Figure~\ref{color_residuals} to actual color changes. These are $0.0062\pm0.0041$\,mag/Gyr for $C_{-3.5}$ and $0.0078\pm0.0026$\,mag/Gyr for $C_{-3.0}$. While this is a very small effect for the age range observed in our globular clusters (10\,Gyr to 14\,Gyr), it can make significant differences when extrapolated to younger populations; e.g. an 8\,Gyr old population would be bluer than predicted by our relation by about 0.03\,mag. 
Note also that the S index does not show any systematic trends with age.

To test for problems with the extinction values, we looked for trends with E(B-V) and galactic latitude (Fig.~\ref{residuals2} and Fig.~\ref{S_residuals2}, middle and lower panel). We do not find any systematics here. 

\subsection{Comparison}
\label{comparison}

We can compare our relations to those derived from stellar evolution models, and to relations from ground-based data transformed to the HST/ACS filter systems.

For comparison with theoretical relations, we use the isochrone set from the Padua, Dartmouth and BaSTI groups. For all isochrone sets we used ages of 8\,Gyr, 10\,Gyr and 13\,Gyr. For Dartmouth, we use $\alpha$-enhancements of $[\alpha/Fe]=\{0.0,0.2,0.4\}$ and for BaSTI models $[\alpha/Fe]=\{0.0,0.4\}$. The PARSEC\footnote{Actually, \citet{bressan12} write about $\alpha$-enhanced PARSEC isochrones, but these are not (yet) publicly available.} and Padua isochrones are available only with solar scaled abundances.
\begin{figure*}[!ht]
\centering
 \begin{minipage}{0.49\textwidth}
 \includegraphics[width=0.99\textwidth]{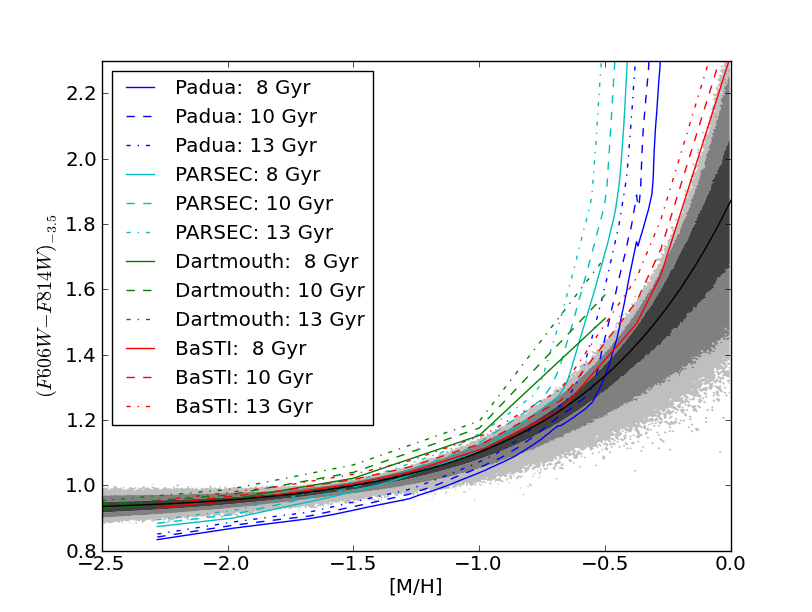}
 \includegraphics[width=0.99\textwidth]{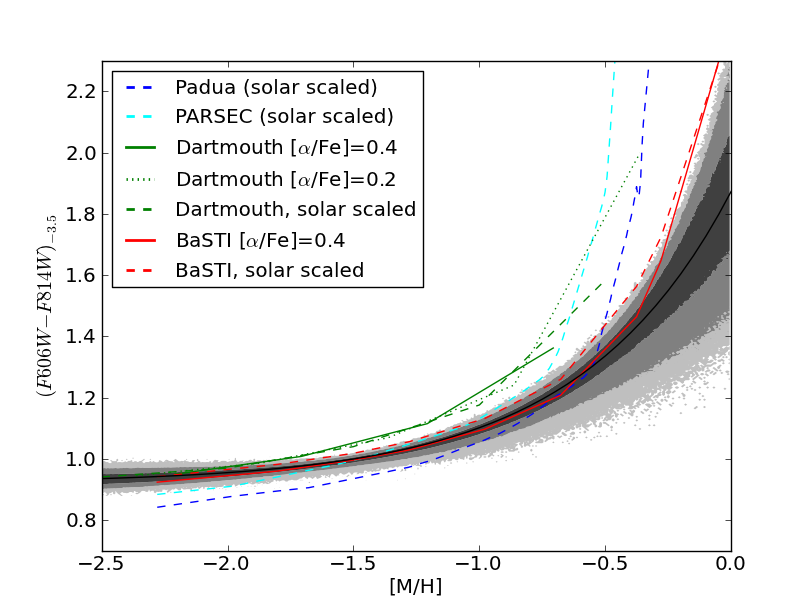}
 \includegraphics[width=0.99\textwidth]{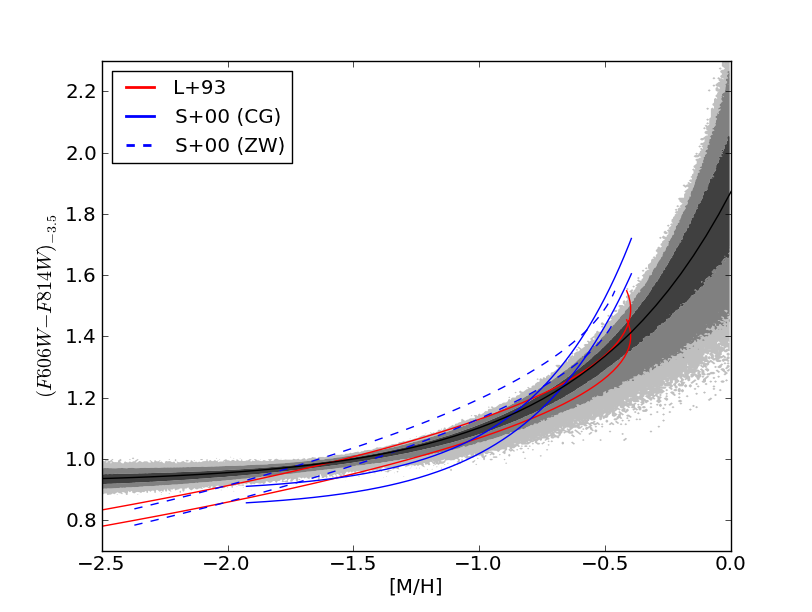}
 \end{minipage}
 \begin{minipage}{0.49\textwidth}
 \includegraphics[width=0.99\textwidth]{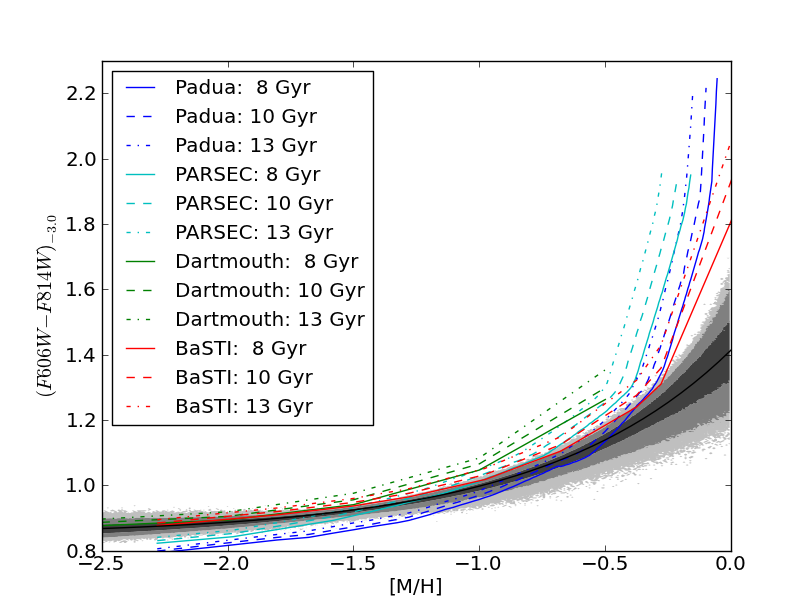}
 \includegraphics[width=0.99\textwidth]{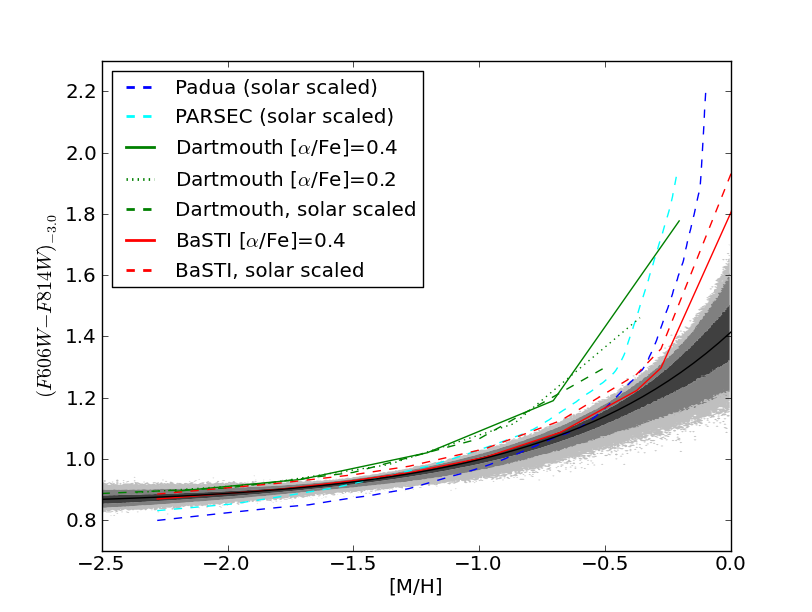}
 \includegraphics[width=0.99\textwidth]{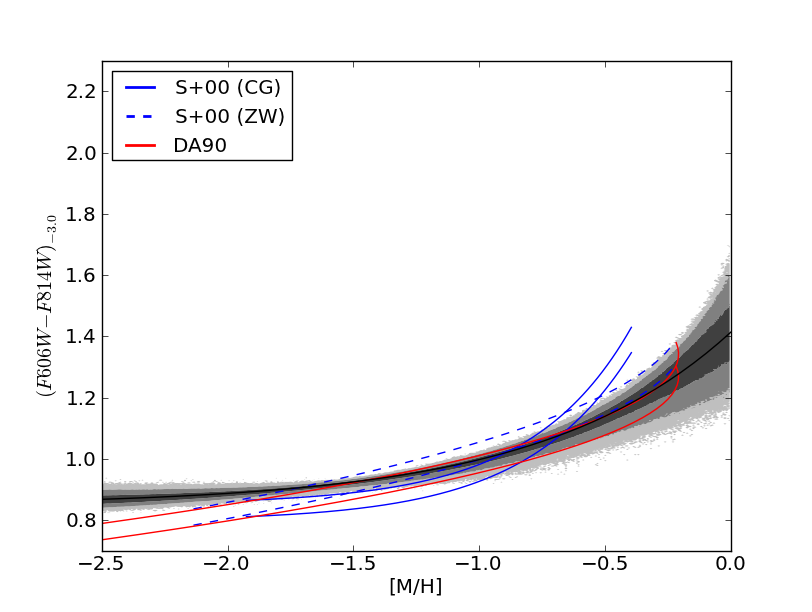}
 \end{minipage}
 \caption{Comparison of color-metallicity relations with theoretical isochrones of different age (top row) and different $\alpha$-enhancement (middle row) and transformed (V-I) relations (bottom row). Left panels compare the relations at $F814W=-3.5$, right panels at $F814W=-3.0$. The (V-I) relations are taken from \citet[][S+00]{saviane00}, \citet[][L+93]{lee93} and \citet[][DA90]{dacosta90}. The S+00 relations are given on two metallicity scales: the ZW scale \citep{zinn84} and the CG scale \citep{carretta97}. The gray contours show the 1$\sigma$, 2$\sigma$, and  3$\sigma$ confidence levels of the fit. The two lines for the observational relations are due to differences between the observational and synthetic transformations from \citet{sirianni05}.}
\label{relations}
\end{figure*}

\begin{figure}[!ht]
 \centering
 \includegraphics[width=0.99\columnwidth]{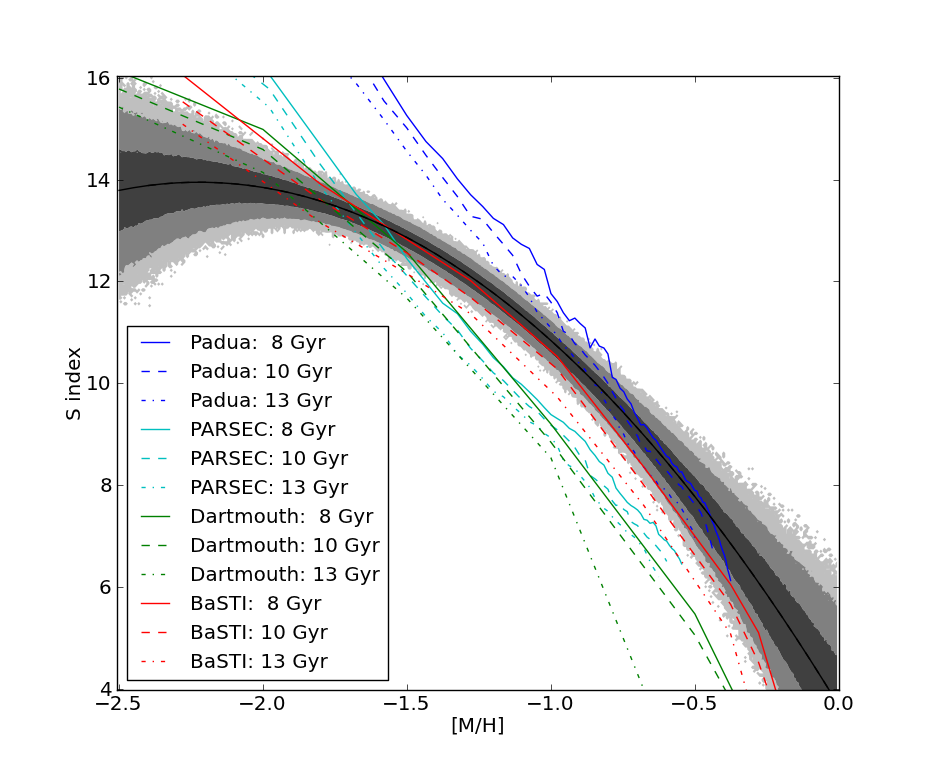}
 \caption{Comparison of the observed S-index metallicity relation with isochrones of varying age. The gray contours show the 1$\sigma$, 2$\sigma$, and  3$\sigma$ confidence levels of the fit.}
\label{relations_sindex}
\end{figure}

All these isochrone sets show a qualitatively similar behavior. The RGB gets redder (Fig.~\ref{relations}) and shallower (Fig.~\ref{relations_sindex}) with increasing metallicity. A higher age also leads to a redder RGB, but this effect is relatively small. An age difference of 5\,Gyr causes the same color difference as a metallicity difference of only 0.1\,dex (see Fig.~\ref{relations}, top row). At a given total metallicity [M/H], the $\alpha$-abundance has almost no effect on the RGB color (Fig.~\ref{relations}, middle row). This supports the assumption that the color of the RGB is mainly influenced by [M/H] and not [Fe/H].

The increasing curvature of the RGB with increasing metallicity prevents the RGB of some metal rich clusters from reaching F814W$=-3.5$\,mag, but bend down at fainter magnitudes. In the Dartmouth models this applies for isochrones with [M/H]$>-0.4$, in Padua models isochrones with [M/H]$>-0.3$. However, the BaSTI RGB isochrones all reach F814W$=-3.5$\,mag, even at super-solar metallicities. Among our clusters, NGC\,6838 ([M/H]$=-0.53$; it also has very few stars in the RGB) and NGC\,6441 ([M/H]$=-0.29$) are affected by this.

In order to quantify the agreement between our relations and other relations, we have performed a Monte Carlo resampling of our relations by drawing random parameter sets $a_i$ from a multivariate Gaussian distribution with the mean and covariance matrix as given by the best fit. The 68.3\%, 95.5\%, and 99.7\% confidence interval are shown as contours in Figures~\ref{metalfitindices}, \ref{s_index}, \ref{relations}, and \ref{relations_sindex}.

As can be seen in Figures~\ref{relations} and~\ref{relations_sindex}, BaSTI isochrones show good agreement with our observational result. At most metallicities the $\alpha$-enhanced BaSTI isochrone falls within the the 1$\sigma$ confidence range of our observational relation; only for [M/H]$>$-0.4 are the isochrones significantly redder ($>3\sigma$) and shallower than our relation. The Dartmouth isochrones agree well at very low metallicities, but tend to predict slightly redder colors and shallower slopes at intermediate and higher metallicities. In contrast, results from the Padua isochrones are bluer by almost 0.15\,mag and much steeper at lower metallicities, and redder and shallower at the high metallicity end. Determining the reason for this offset is beyond the scope of this paper, but this problem has been known to lead to higher metallicity estimates, when Padua isochrones are used \citep{lejeune99}.

\paragraph{}
Existing color-metallicity relations are given in the standard Johnson-Cousin filters. Thus to compare these with our analysis we use the transformations to the HST/ACS filter set described in \citet{sirianni05}. Two such transformations are provided, one observationally based and the other synthetic. The former uses observations of horizontal branch and RGB stars in the metal-poor ([Fe/H]=$-2.15$) globular cluster NGC\,2419. This cluster does not contain stars with $(V-I) > 1.3$, hence the transformation at these redder colors are extrapolated and should be used with caution. The transformation is:

$$ F606W - F814W = -0.055 + 0.762(V - I)$$

For the synthetic transformation, stellar models with $(V-I) < 1.8$ were used. Hence, for redder colors, the extrapolation should again be treated with caution. The transformation is given as:

$$F606W - F814W = 0.062 + 0.646(V - I) + 0.053(V - I)^2$$

We use both these transformations on the color-metallicity relations of \citet{saviane00}, who determined relations for the indices $(V - I)_{-3.0}$ and $(V - I)_{-3.5}$, and for \citet[][for $(V - I)_{-3.0}$]{dacosta90} and \citet[][for $(V - I)_{-3.5}$]{lee93}.
To shift these transformations, which are defined for [Fe/H], to the [M/H] scale, we used the same [Fe/H]-[$\alpha$/Fe] relation as for the data.

Note that these two transformations have a relative offset of of about 0.05\,mag, which can be seen in Fig.~\ref{relations} as the two almost parallel lines in the lower panel.\footnote{The offset can already be seen in \citet[][Fig.~21]{sirianni05} as an offset in plot of (V-I) versus V-F606W.}

From Fig.~\ref{relations} it can be seen that these transformed relations are always bluer at the low metallicity end and have a steeper slope than our relations.\footnote{Strictly speaking, we compare slightly different things here: The transformed relations measure the color at constant I-band magnitude, while in this work we have measured the color at constant F814W magnitude. We can ignore this difference here because the difference between I-band and F814W is small. According to the transformations given above, the differences between F814W and I are always smaller than 0.05\,mag and the resulting error in the color measurement of the RGB is always smaller than 0.01\,mag (except for the two reddest clusters, for which it can reach 0.06~mag). Therefore the effect on the total color-metallicity relation is negligible.} 

\begin{figure}[!ht]
 \centering
 \includegraphics[width=0.99\columnwidth]{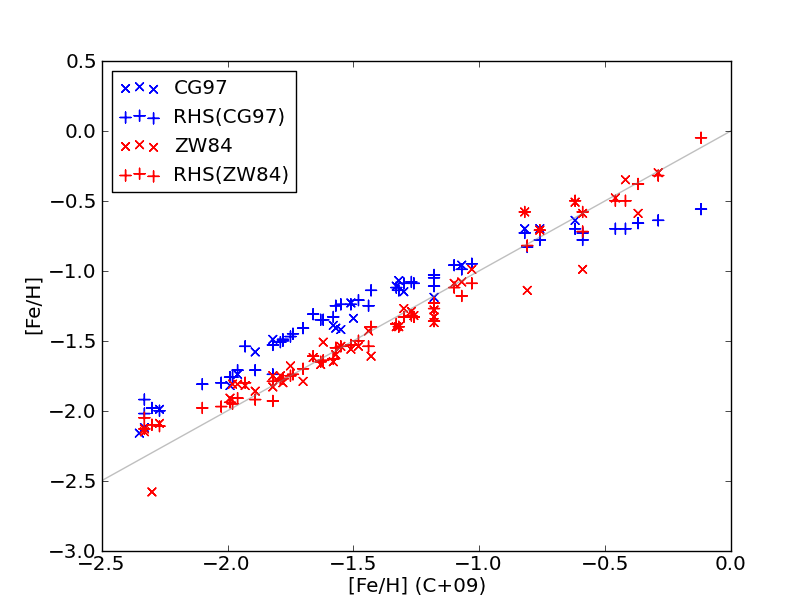}
 \caption[Comparison of metallicity scales]{Comparison of different metallicity scales. On the x-axis the C+09 scale, that is adopted in this paper, is shown. Blue crosses show the clusters from \citet{carretta97}, red crosses from \citet{zinn84}. The plus symbols show the metallicities determined in \citet[][RHS]{rutledge97} based on the Ca triplet, calibrated to both scales.}
\label{fehscales}
\end{figure}

Part of this discrepancy can be explained by the different metallicity scales used for the various relations. While we use the metallicity scale of \citet[][,C+09]{carretta09}, earlier relations were determined either in the Zinn \& West scale \citep[][ZW84]{zinn84} or the Carretta \& Gratton scale \citep[][CG97]{carretta97}. The adopted C+09 scale is comparable to the ZW84 scale; however, the CG97 scale yields higher metallicities for [Fe/H]$\lesssim-1$ and lower metallicities for [Fe/H]$\gtrsim-1$ (see~\ref{fehscales}).

Hence, using the CG97 scale will lead to a steeper color-metallicty relation than found from our measurements (see the bottom panel of Fig.~\ref{relations}).

\subsection{Inverting the relation}
\begin{figure}[!ht]
\centering
 \includegraphics[width=0.99\columnwidth]{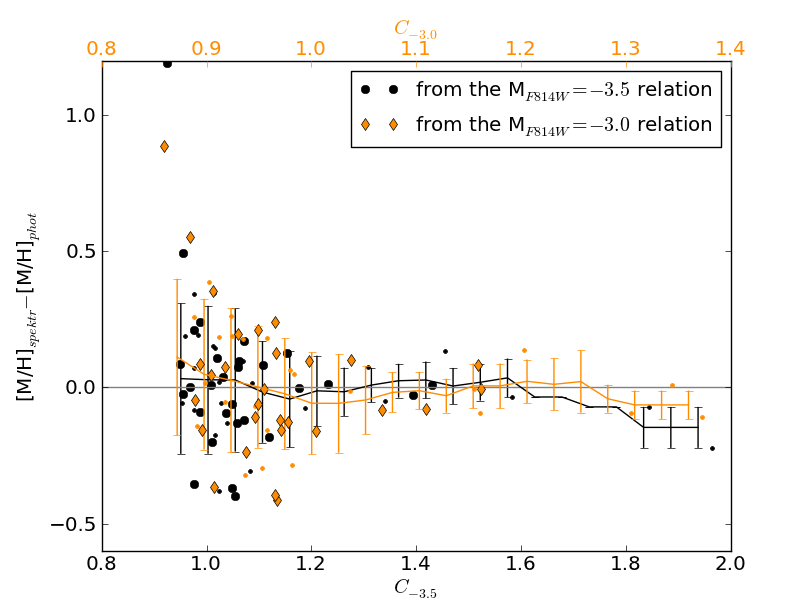}
 \caption{Error distribution of the metallicity determination using the inverted color metallicity relations. Lines with errorbars are the running mean and standard deviation which are computed using a bin width of 0.3\,mag for $C_{-3.5}$ and 0.15\,mag for $C_{-3.0}$. Symbols and colors are as in Fig.~\ref{metalfitindices}. Note the different scales on the x-axis for the two distributions.}
\label{inverted_residuals}
\end{figure}

The main purpose of the color metallicity relation is to estimate metallicities of old stellar population. The uncertainties arising from the inverted relation are highly nonlinear. In Fig.~\ref{inverted_residuals} we plot the difference between the spectroscopic metallicities and the metallicities derived with our relation. It is apparent that for bluer colors (i.e. lower metallicities) the difference can be very large. If the color is near the pole of the metallicity-color function, the formal uncertainties can be infinite. Then only an upper limit on the metallicity can be derived. For all clusters with $C_{-3.5}<1.2$ (or $C_{-3.0}<1.0$) the scatter in the metallicity differences is about 0.3\,dex. We suggest using this as a minimum uncertainty for metallicities derived from our relation in that color range. For redder colors, the uncertainty drops in half.

\section{Conclusions and summary}
\label{summary}

In this paper, we derived relations between the colors and the slope of the RGB and metallicity using data from globular clusters. The details of the relations are summarized in Table~\ref{fitparams}.
When using these relations for determining metallicities of old resolved stellar populations, the following points should be kept in mind:
\begin{itemize}
\item The color changes very little with metallicity for $\mbox{[M/H]}\lesssim -1.0$, the slope changes little below $[M/H]\lesssim-1.5$. Therefore, inverting the relation in this regime introduces large uncertainties. This makes a photometric metallicity determination rather inaccurate in this metallicity range.
\item Our relation agrees well with the prediction from BaSTI isochrones. Dartmouth isochrones are slightly redder, Padua isochrones bluer than our data. Thus, for the purpose of determining metallicities of old populations we recommend the use of BaSTI isochrones.
\item A comparison with other color-metallicity-relations from the literature, both empirical and theoretical, shows some scatter between these relations. Therefore a comparison of metallicities derived from different methods/relations will introduce systematic offsets. This should be kept in mind whenever the use of a homogenous method is not possible.
\end{itemize}

\begin{acknowledgements}
We thank our referee, Ivo Saviane, for the careful reading of our manuscript, the detailed look into the data, and the comments that helped to improve this paper.
We also thank Benne Holwerda and Antonela Monachesi for their comments on an earlier version of this paper that improved its final quality.

DS gratefully acknowledges the support from DLR via grant 50OR1012 and a scholarship from the Cusanuswerk.
\end{acknowledgements}

\bibliographystyle{aa}
\bibliography{../../bibliography}

\appendix

\section{Description of the fit of the RGB}

\begin{figure*}[!ht]
\centering
 \includegraphics[width=0.99\textwidth]{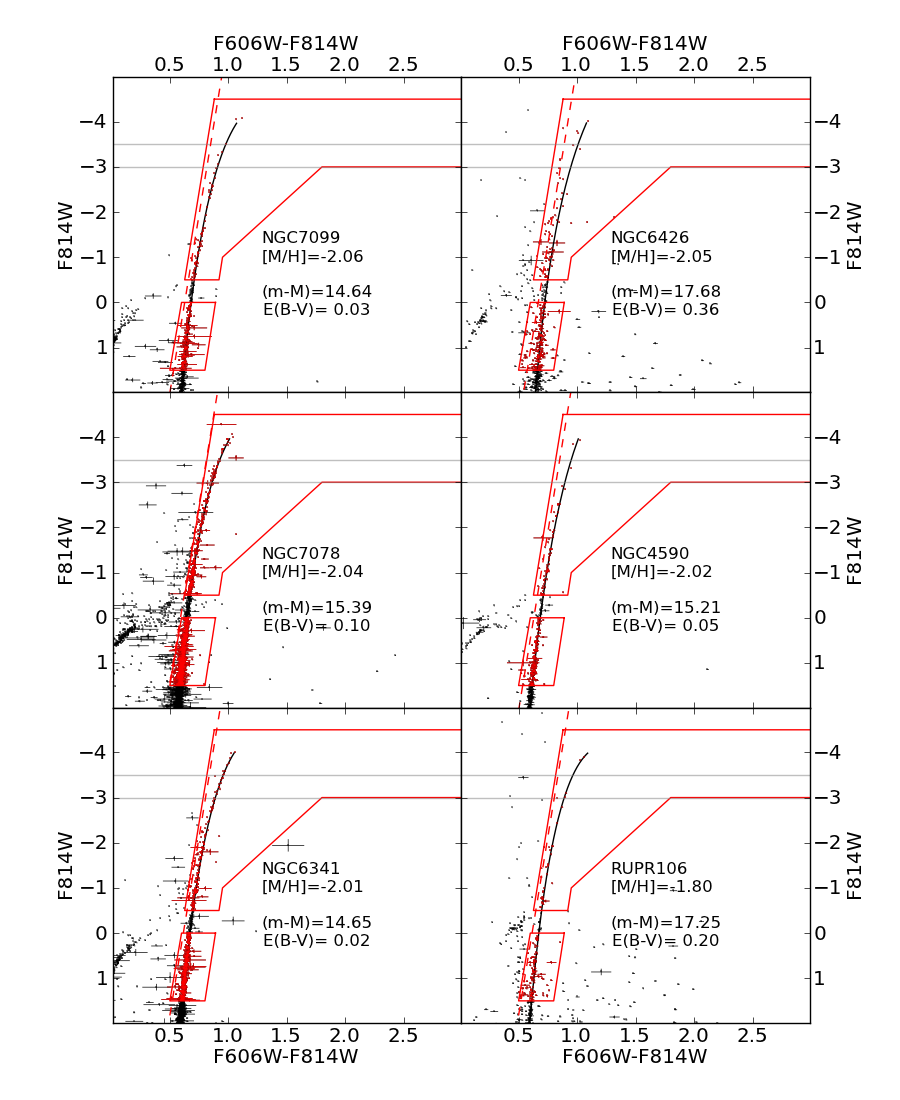}
 \caption[]{CMDs of the 53 clusters in our sample, with fitted RGB curves. The two horizontal gray lines indicate the magnitudes at which the colors are measured. The red frame gives the region where stars are used for the fit. The reference line to separate AGB from RGB stars is shown as red dashed line. The solid black line shows the best-fit RGB function after the rejection of the AGB. Note, that all stars in the CMD are actually drawn with errorbars, but for many stars the reported errors are too small to be visible in these plots.}
 \label{CMD_diverse}
\end{figure*}

We parametrized the RGB with the function 
$$ M=a+b\cdot \mbox{color}+c/(\mbox{color}+d) ,$$
as given in \citet{saviane00}. Since the data do not only contain RGB stars, but also the horizontal branch, blue stragglers and foreground stars, we have defined a region to guarantee a high fraction of RGB stars in our fit sample. The extent of this region can be seen as the red frame in Fig.~\ref{CMD_diverse}. Note that this region excludes also the red clump.

In some clusters, there is a distinct asymptotic giant branch (AGB) visible, which lies on the blue side of the RGB (it is mostly seen at F814W magnitudes between -1 and -2). 
As in \citet{saviane00}, we have removed these AGB stars by excluding all detections that lie blue wards of a reference line with the same slope for all clusters (denoted in Figures~\ref{CMD_diverse} through~\ref{CMDs_diverse_last} by a dashed red line). The horizontal position of the reference line was set to be 0.05\,mag blue wards (at F814W=-0.5) of a first fit of all stars in the RGB region and then excluding. The fit including all stars is shown in the CMDs as black dashed line, while the final fit after the AGB removal is shown as black solid line.

For the actual fit we used the python package scipy.odr. This routine performs an orthogonal distance regression, i.e. it minimizes the orthogonal distance between the curve and the data points. The distance of each data point is weighted with its measurement uncertainty. This method is a variation of the typical $\chi^2$ minimization, now generalized for data with uncertainties on both variables.

The ACSGCS team reports photometric uncertainties for each individual star, which are typically quite small; the median uncertainty in F814W is only 0.003\,mag. This is much smaller than both the observed scatter in the RGB and the errors that are found in the artificial star test at a level of F814W$\approx0$. (There are no artificial star tests at brighter magnitudes.) The mean measurement error estimated from the difference of the input and recovered magnitudes in the artificial star test are 0.06\,mag in F814W and 0.03\,mag in color. These estimated are added in quadrature to the reported uncertainties of each star. 
The smaller error in color is due to the fact that errors in both bands are correlated. Thus the uncertainty of the difference of both bands is smaller than the uncertainty in each band.

Finally, we visually inspected each CMD with its fit, to check for any residual problems of our clusters. After this inspection we excluded four more clusters from the sample: NGC\,6838 and NGC\,6441, because their RGB fits do not reach the $M_I=-3.5$ level; NGC\,6388 (and again NGC\,6441), because their red clumps seem to be to faint \citep[][also found problems with differential reddening and multiple populations in these two clusters]{bellini13}; and NGC\,6715, because it has a clear and strong second RGB.

\begin{figure*}[!ht]
\centering
 \includegraphics[width=0.99\textwidth]{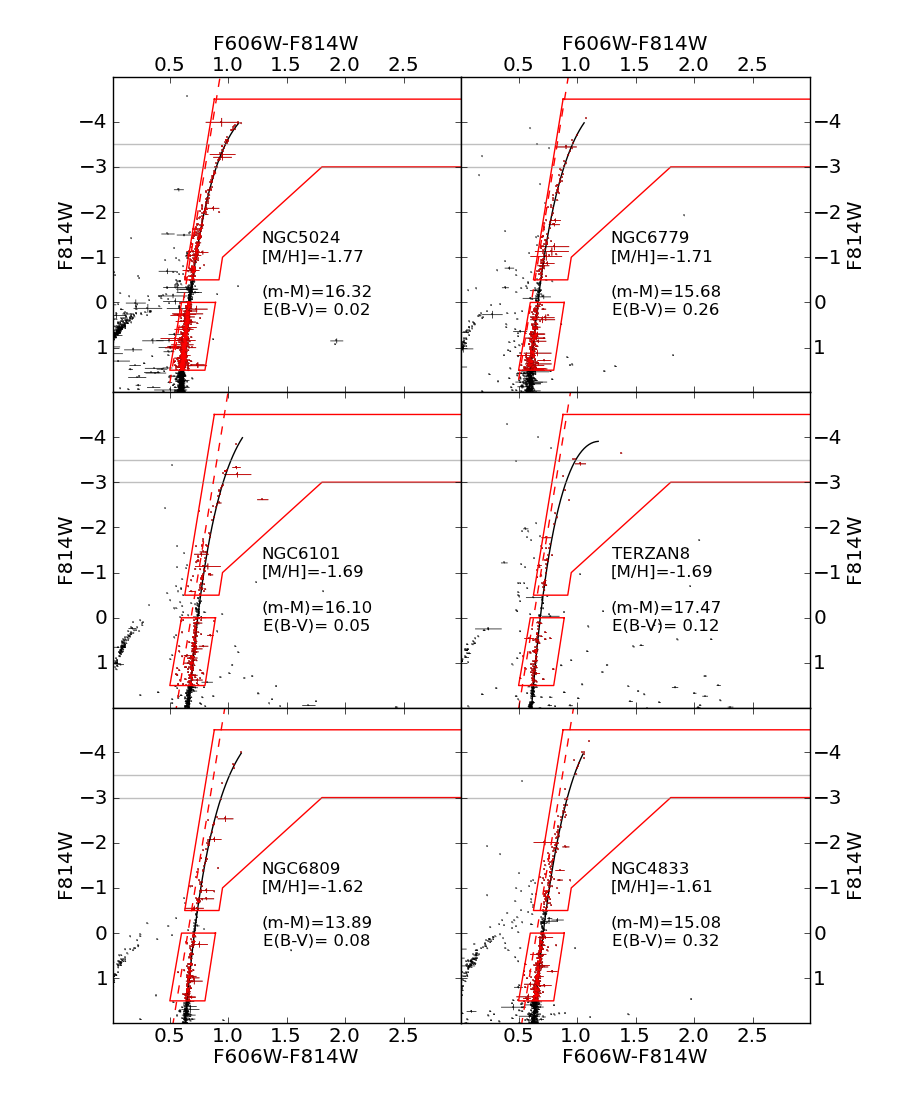}
 \caption{as Fig.~\ref{CMD_diverse}.}
\end{figure*}
\begin{figure*}[!ht]
\centering
 \includegraphics[width=0.99\textwidth]{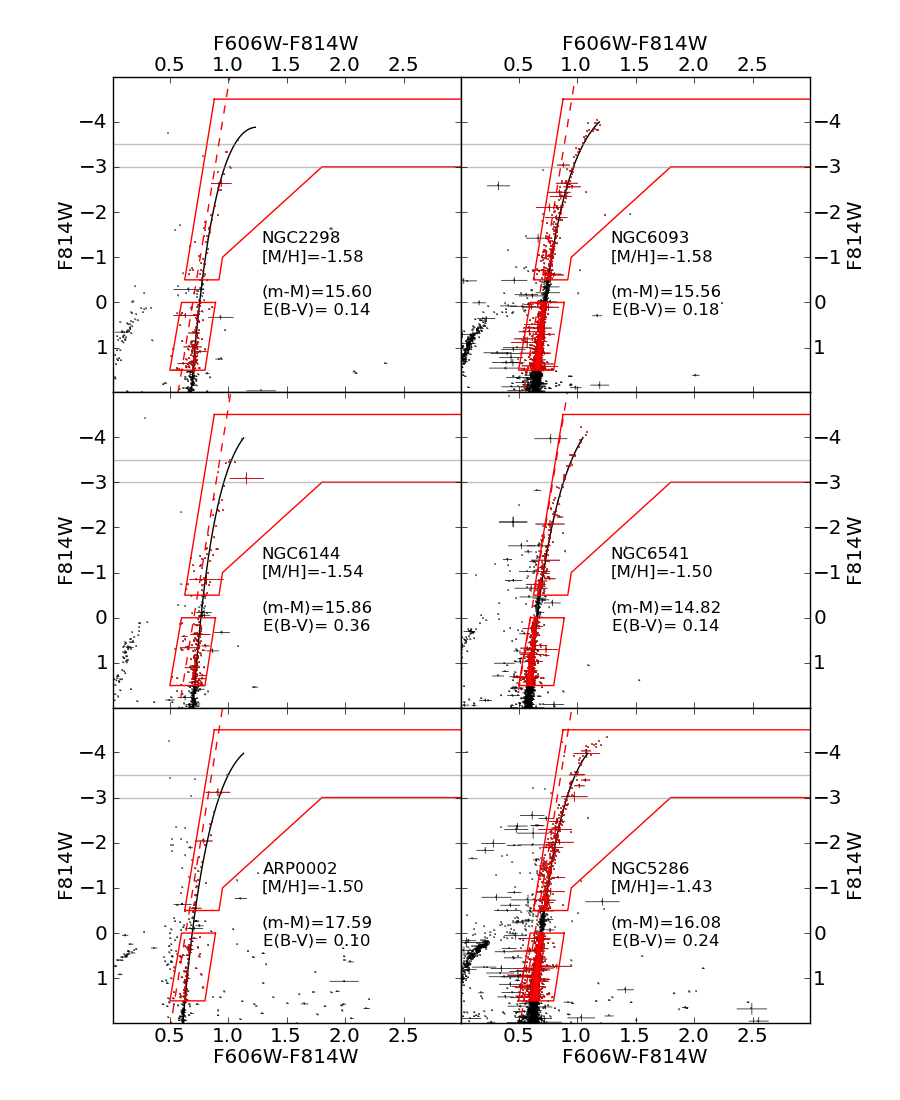}
 \caption{as Fig.~\ref{CMD_diverse}.}
\end{figure*}
\begin{figure*}[!ht]
\centering
 \includegraphics[width=0.99\textwidth]{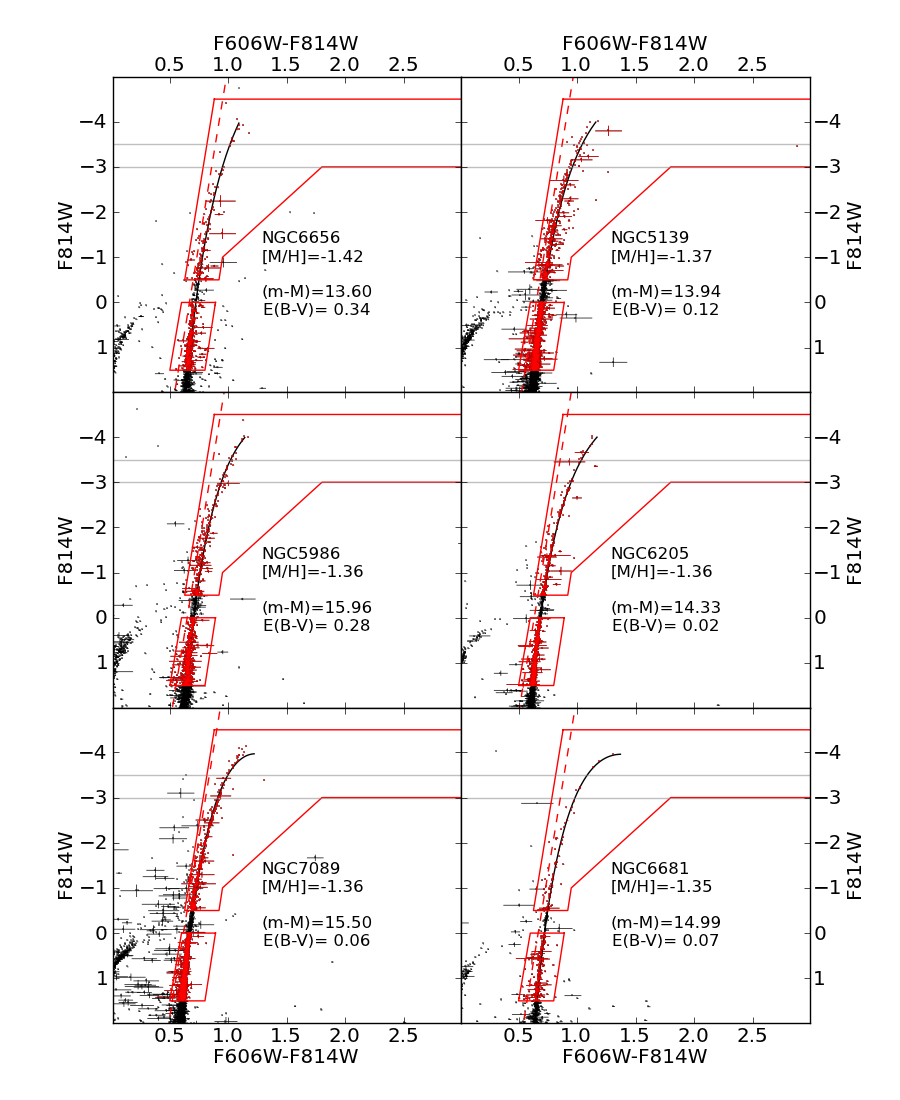}
 \caption{as Fig.~\ref{CMD_diverse}.}
\end{figure*}
\begin{figure*}[!ht]
\centering
 \includegraphics[width=0.99\textwidth]{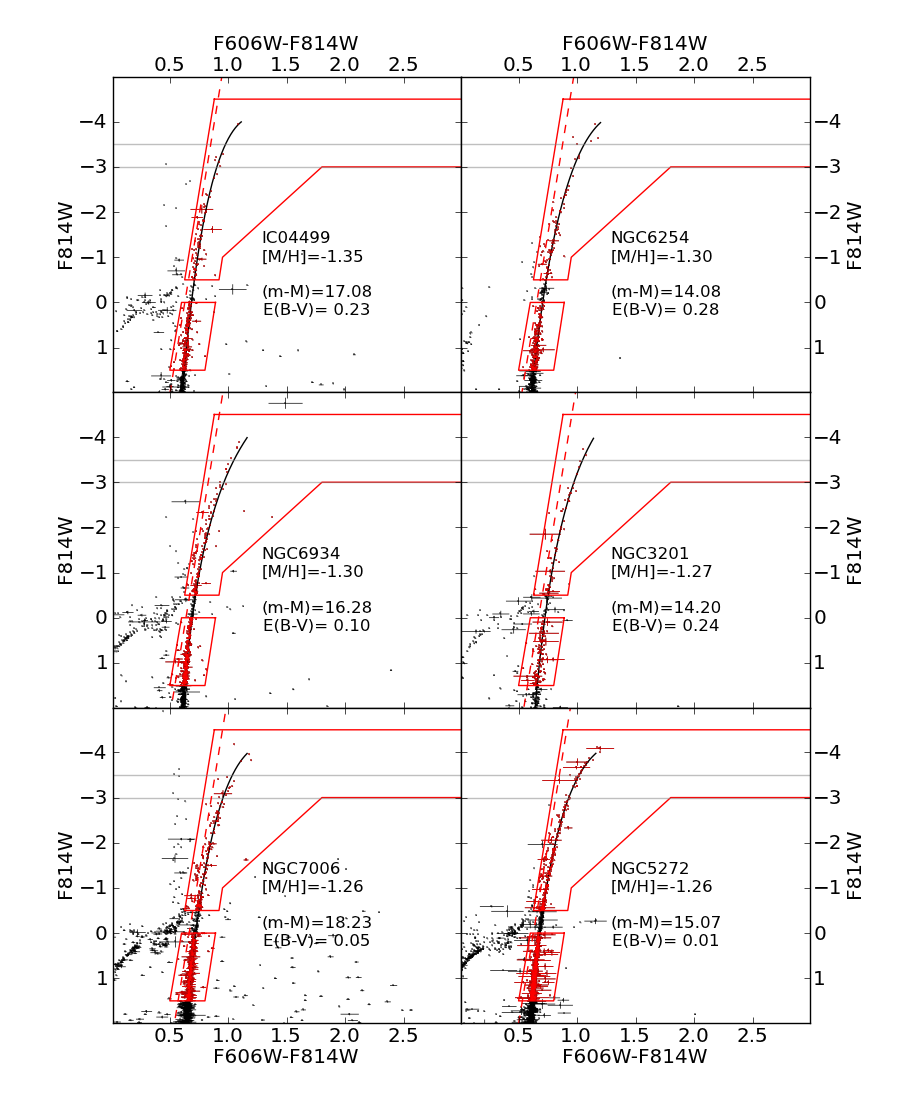}
 \caption{as Fig.~\ref{CMD_diverse}.}
\end{figure*}
\begin{figure*}[!ht]
\centering
 \includegraphics[width=0.99\textwidth]{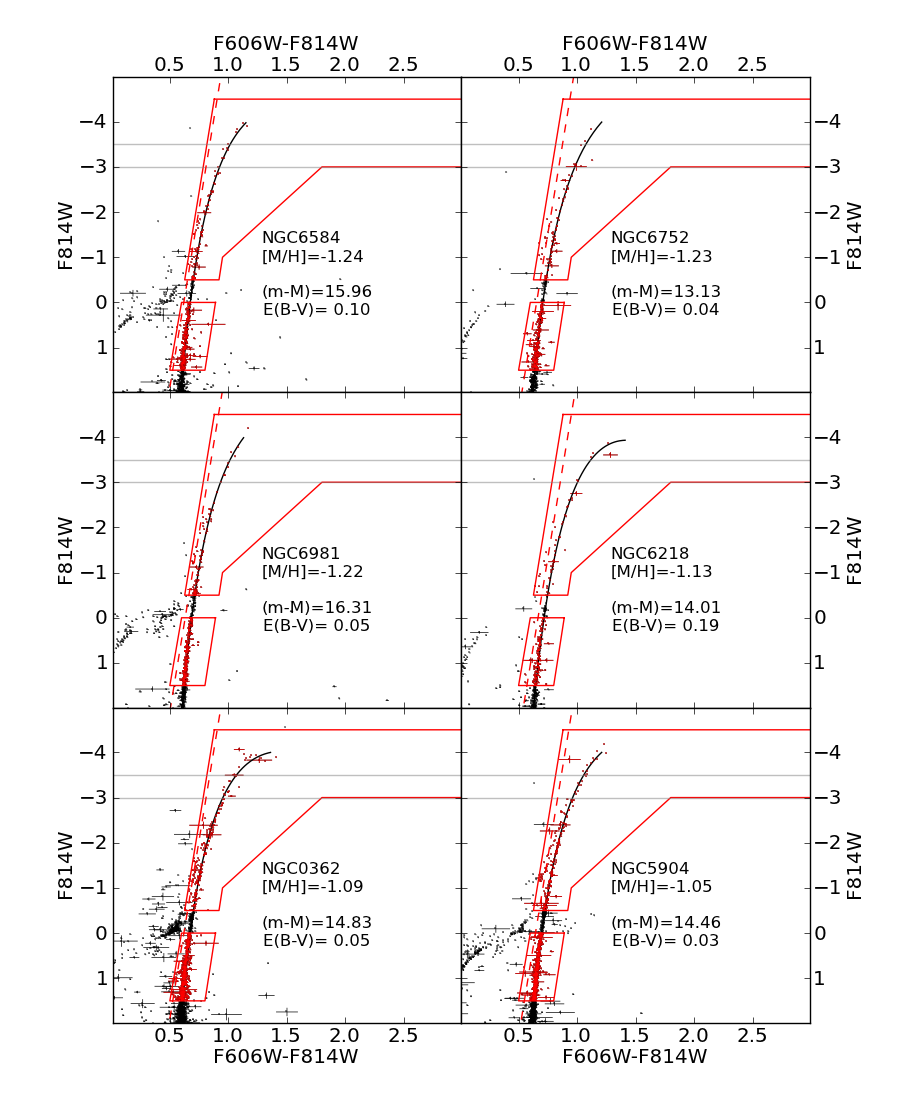}
 \caption{as Fig.~\ref{CMD_diverse}.}
\end{figure*}
\begin{figure*}[!ht]
\centering
 \includegraphics[width=0.99\textwidth]{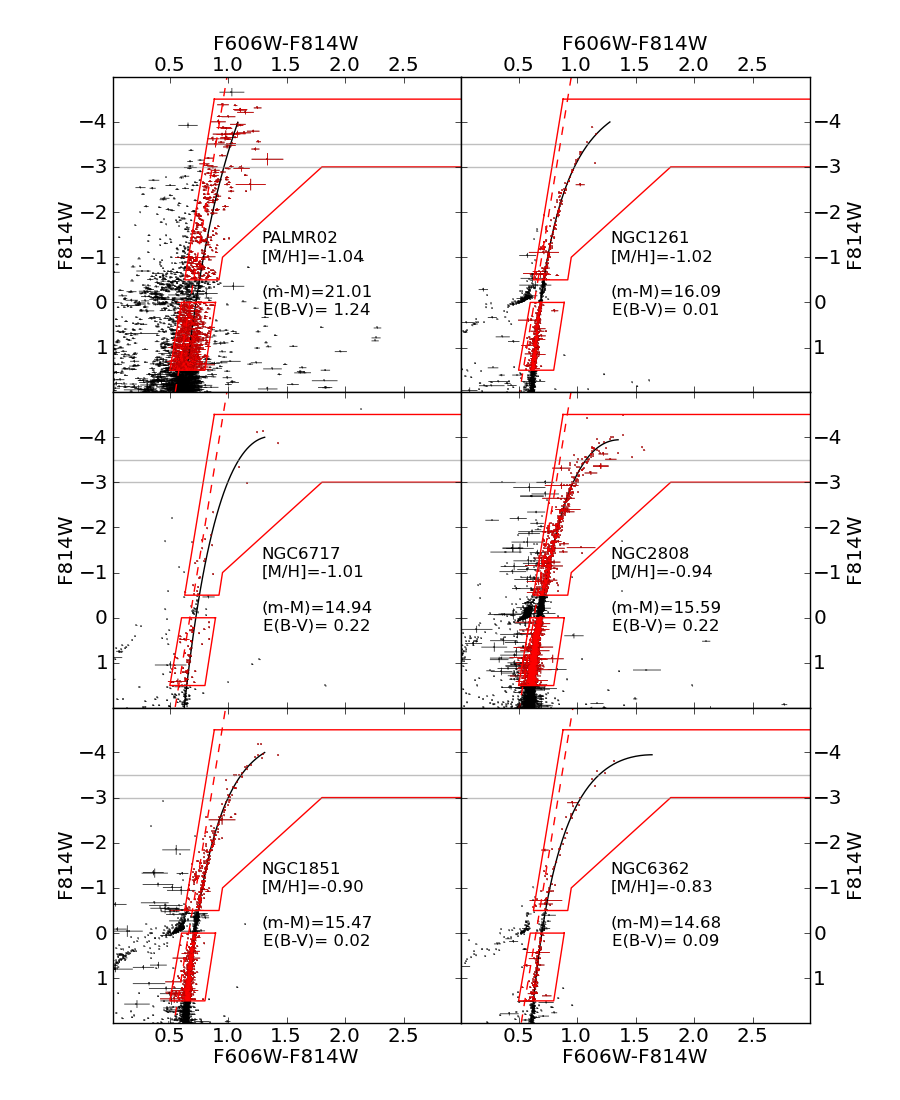}
 \caption{as Fig.~\ref{CMD_diverse}.}
\end{figure*}
\begin{figure*}[!ht]
\centering
 \includegraphics[width=0.99\textwidth]{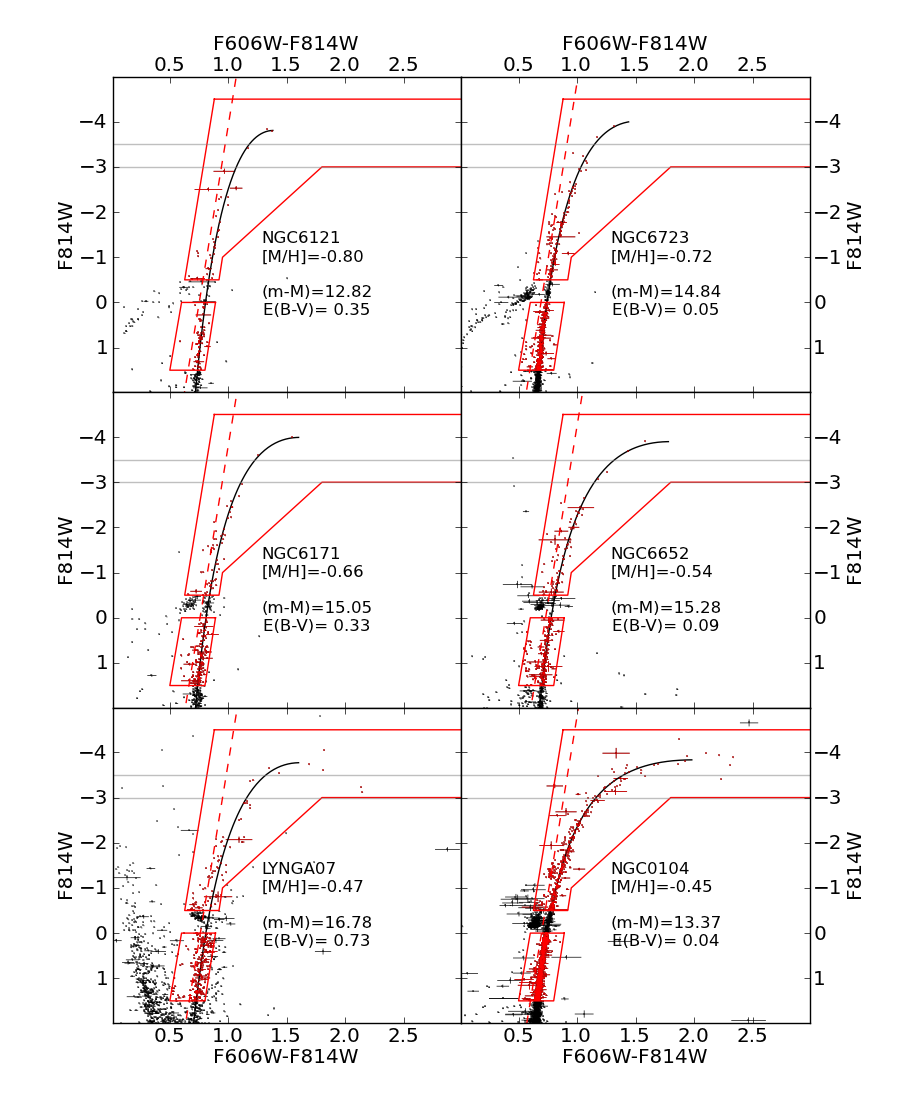}
 \caption{as Fig.~\ref{CMD_diverse}.}
\end{figure*}
\begin{figure*}[!ht]
\centering
 \includegraphics[width=0.99\textwidth]{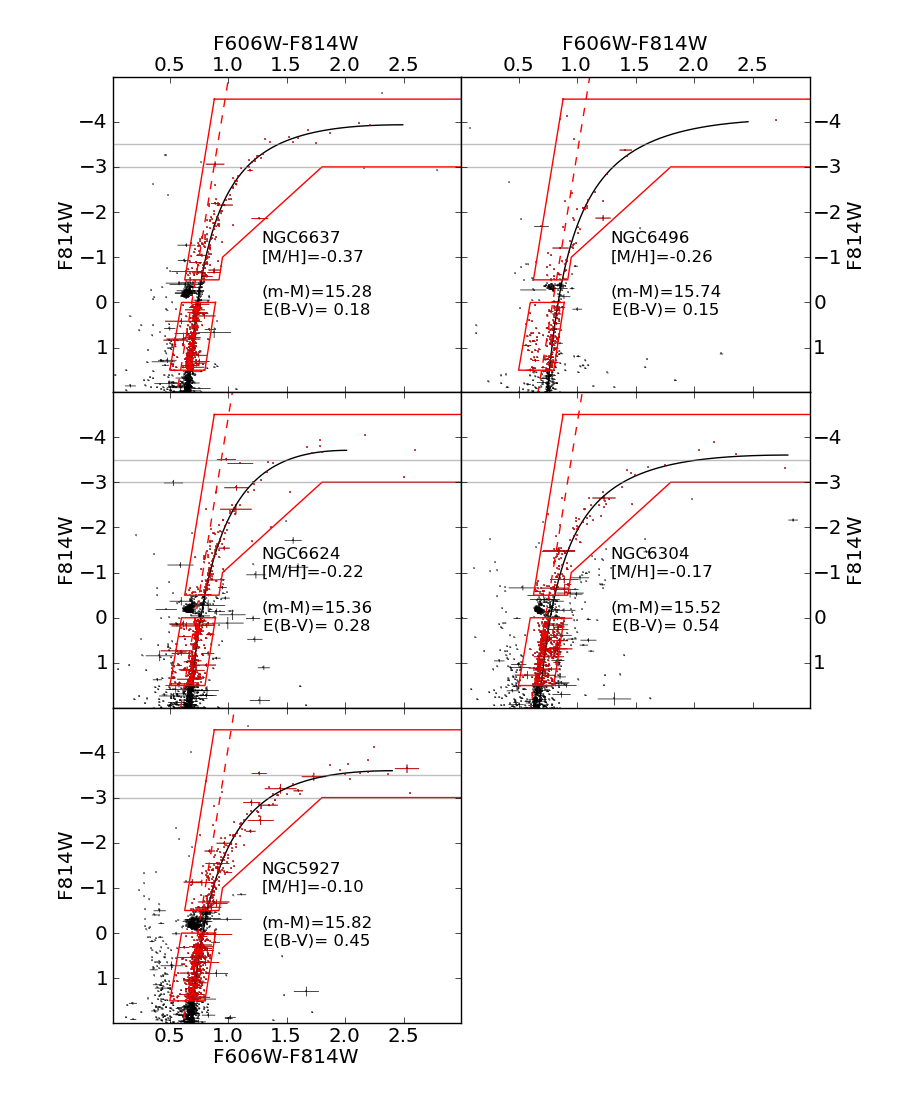}
 \caption{as Fig.~\ref{CMD_diverse}.}
\label{CMDs_diverse_last}
\end{figure*}

\section{Properties of the Globular clusters} \label{app:properties}
The properties of the globular clusters, from the literature and determined in this work, are summarized in Table~\ref{GC_literature} and Table~\ref{GC_results} respectively.
\begin{table*}[!ht]
\centering
\caption{Literature data for the clusters used in this work.}
\begin{tabular}{c|ccccccccc}
name & RA & DEC & (m-M$)_V$ & E(B-V) & [Fe/H]$_{H10}$ & [Fe/H]$_{C+10}$ & $\sigma_{[Fe/H]}$ & [$\alpha$/Fe] & age \\
     & [$\deg$] & [$\deg$] & [mag] & [mag] & [dex] & [dex] & [dex] & [dex] & \\
 (1) & (2) & (3) & (4) & (5) & (6) & (7) & (8) & (9) & (10) \\ \hline
Arp~2     & 292.1838 & -29.6444 &    17.59 &   0.10 &  -1.75 &  -1.74 &   0.08 &   0.34 &   0.85 \\
IC\,4499  & 225.0769 & -81.7863 &    17.08 &   0.23 &  -1.53 &  -1.62 &   0.09 &   ...  &   ...  \\
Lynga~7   & 242.7652 & -54.6822 &    16.78 &   0.73 &  -1.01 &  -0.68 &   0.06 &   ...  &   1.13 \\
NGC\,104  &   6.0236 & -71.9187 &    13.37 &   0.04 &  -0.72 &  -0.76 &   0.02 &   0.42 &   1.02 \\
NGC\,362  &  15.8094 & -69.1512 &    14.83 &   0.05 &  -1.26 &  -1.30 &   0.04 &   0.30 &   0.81 \\
NGC\,1261 &  48.0675 & -54.7838 &    16.09 &   0.01 &  -1.27 &  -1.27 &   0.08 &   ...  &   0.80 \\
NGC\,1851 &  78.5282 & -39.9534 &    15.47 &   0.02 &  -1.18 &  -1.18 &   0.08 &   0.38 &   0.78 \\
NGC\,2298 & 102.2475 & -35.9947 &    15.60 &   0.14 &  -1.92 &  -1.96 &   0.04 &   0.50 &   0.99 \\
NGC\,2808 & 138.0129 & -63.1365 &    15.59 &   0.22 &  -1.14 &  -1.18 &   0.04 &   0.33 &   0.85 \\
NGC\,3201 & 154.4034 & -45.5875 &    14.20 &   0.24 &  -1.59 &  -1.51 &   0.02 &   0.33 &   0.80 \\
NGC\,4590 & 189.8666 & -25.2559 &    15.21 &   0.05 &  -2.23 &  -2.27 &   0.04 &   0.35 &   0.90 \\
NGC\,4833 & 194.8913 & -69.1235 &    15.08 &   0.32 &  -1.85 &  -1.89 &   0.05 &   ...  &   0.98 \\
NGC\,5024 & 198.2302 &  18.1682 &    16.32 &   0.02 &  -2.10 &  -2.06 &   0.09 &   ...  &   0.99 \\
NGC\,5139 & 201.6968 & -46.5204 &    13.94 &   0.12 &  -1.53 &  -1.64 &   0.09 &   ...  &   0.90 \\
NGC\,5272 & 205.5484 &  28.3773 &    15.07 &   0.01 &  -1.50 &  -1.50 &   0.05 &   0.34 &   0.89 \\
NGC\,5286 & 206.6117 & -50.6257 &    16.08 &   0.24 &  -1.69 &  -1.70 &   0.07 &   ...  &   0.98 \\
NGC\,5904 & 229.6384 &   2.0810 &    14.46 &   0.03 &  -1.29 &  -1.33 &   0.02 &   0.38 &   0.83 \\
NGC\,5927 & 232.0029 & -49.3270 &    15.82 &   0.45 &  -0.49 &  -0.29 &   0.07 &   ...  &   0.99 \\
NGC\,5986 & 236.5125 & -36.2136 &    15.96 &   0.28 &  -1.59 &  -1.63 &   0.08 &   ...  &   0.95 \\
NGC\,6093 & 244.2600 & -21.0239 &    15.56 &   0.18 &  -1.75 &  -1.75 &   0.08 &   0.24 &   0.98 \\
NGC\,6101 & 246.4505 & -71.7978 &    16.10 &   0.05 &  -1.98 &  -1.98 &   0.07 &   ...  &   0.98 \\
NGC\,6121 & 245.8968 & -25.4743 &    12.82 &   0.35 &  -1.16 &  -1.18 &   0.02 &   0.51 &   0.98 \\
NGC\,6144 & 246.8077 & -25.9765 &    15.86 &   0.36 &  -1.76 &  -1.82 &   0.05 &   ...  &   1.08 \\
NGC\,6171 & 248.1328 & -12.9462 &    15.05 &   0.33 &  -1.02 &  -1.03 &   0.02 &   0.49 &   1.09 \\
NGC\,6205 & 250.4218 &  36.4599 &    14.33 &   0.02 &  -1.53 &  -1.58 &   0.04 &   0.31 &   0.91 \\
NGC\,6218 & 251.8091 &  -0.0515 &    14.01 &   0.19 &  -1.37 &  -1.43 &   0.02 &   0.41 &   0.99 \\
NGC\,6254 & 254.2877 &  -3.8997 &    14.08 &   0.28 &  -1.56 &  -1.57 &   0.02 &   0.37 &   0.89 \\
NGC\,6304 & 258.6344 & -28.5380 &    15.52 &   0.54 &  -0.45 &  -0.37 &   0.07 &   ...  &   1.06 \\
NGC\,6341 & 259.2808 &  43.1359 &    14.65 &   0.02 &  -2.31 &  -2.35 &   0.05 &   0.46 &   1.03 \\
NGC\,6362 & 262.9791 & -66.9517 &    14.68 &   0.09 &  -0.99 &  -1.07 &   0.05 &   ...  &   1.06 \\
NGC\,6426 & 266.2277 &   3.1701 &    17.68 &   0.36 &  -2.15 &  -2.36 &   0.06 &   ...  &   ...  \\
NGC\,6496 & 269.7653 & -43.7341 &    15.74 &   0.15 &  -0.46 &  -0.46 &   0.07 &   ...  &   0.97 \\
NGC\,6541 & 272.0098 & -42.2851 &    14.82 &   0.14 &  -1.81 &  -1.82 &   0.08 &   0.43 &   1.01 \\
NGC\,6584 & 274.6567 & -51.7842 &    15.96 &   0.10 &  -1.50 &  -1.50 &   0.09 &   ...  &   0.88 \\
NGC\,6624 & 275.9188 & -29.6390 &    15.36 &   0.28 &  -0.44 &  -0.42 &   0.07 &   ...  &   0.98 \\
NGC\,6637 & 277.8462 & -31.6519 &    15.28 &   0.18 &  -0.64 &  -0.59 &   0.07 &   0.31 &   1.02 \\
NGC\,6652 & 278.9401 & -31.0093 &    15.28 &   0.09 &  -0.81 &  -0.76 &   0.14 &   ...  &   1.01 \\
NGC\,6656 & 279.0998 & -22.0953 &    13.60 &   0.34 &  -1.70 &  -1.70 &   0.08 &   0.38 &   0.99 \\
NGC\,6681 & 280.8032 & -31.7079 &    14.99 &   0.07 &  -1.62 &  -1.62 &   0.08 &   ...  &   1.00 \\
NGC\,6717 & 283.7752 & -21.2985 &    14.94 &   0.22 &  -1.26 &  -1.26 &   0.07 &   ...  &   1.03 \\
NGC\,6723 & 284.8881 & -35.3678 &    14.84 &   0.05 &  -1.10 &  -1.10 &   0.07 &   0.50 &   1.02 \\
NGC\,6752 & 287.7171 & -58.0154 &    13.13 &   0.04 &  -1.54 &  -1.55 &   0.01 &   0.43 &   0.92 \\
NGC\,6779 & 289.1482 &  30.1835 &    15.68 &   0.26 &  -1.98 &  -2.00 &   0.09 &   ...  &   1.07 \\
NGC\,6809 & 294.9988 & -29.0353 &    13.89 &   0.08 &  -1.94 &  -1.93 &   0.02 &   0.42 &   0.96 \\
NGC\,6934 & 308.5474 &   7.4045 &    16.28 &   0.10 &  -1.47 &  -1.56 &   0.09 &   ...  &   0.87 \\
NGC\,6981 & 313.3654 & -11.4627 &    16.31 &   0.05 &  -1.42 &  -1.48 &   0.07 &   ...  &   0.85 \\
NGC\,7006 & 315.3724 &  16.1873 &    18.23 &   0.05 &  -1.52 &  -1.46 &   0.06 &   0.28 &   ...  \\
NGC\,7078 & 322.4930 &  12.1670 &    15.39 &   0.10 &  -2.37 &  -2.33 &   0.02 &   0.40 &   1.01 \\
NGC\,7089 & 323.3626 &   0.8233 &    15.50 &   0.06 &  -1.65 &  -1.66 &   0.07 &   0.41 &   0.92 \\
NGC\,7099 & 325.0922 & -22.8201 &    14.64 &   0.03 &  -2.27 &  -2.33 &   0.02 &   0.37 &   1.01 \\
Pal~2     &  71.5246 &  31.3815 &    21.01 &   1.24 &  -1.42 &  -1.29 &   0.09 &   ...  &   ...  \\
Rup~106   & 189.6675 & -50.8497 &    17.25 &   0.20 &  -1.68 &  -1.78 &   0.08 &  -0.03 &   ...  \\
Terzan~8  & 295.4350 & -32.0005 &    17.47 &   0.12 &  -2.16 &  -2.02 &   0.06 &   0.45 &   0.95 \\
\end{tabular}
 \tablefoot{Keys to columns: (1)-(6) name, right ascension, declination, visual distance modulus, color excess and metallicity from \citet{harris10}; (7-9) metallicity, its uncertainty and alpha enhancement from \citet{carretta10}; (10) relative age from \citet{marinfranch09}.}
\label{GC_literature}
\end{table*}

\begin{table*}[!ht]
\centering
\caption{Results for all clusters used in this work.} 
\begin{tabular}{c|cccc|ccc}
name &  (V-I$)_{-3.5}$ & $\sigma_{-3.5}$ & (V-I$)_{-3.0}$ & $\sigma_{-3.0}$ & $S$ & $\sigma_S$ & $M_{V}(HB)$ \\ 
     & [mag] & [mag] & [mag] & [mag] &   &  & [mag] \\
 (1) & (2) & (3) & (4) & (5) & (6) & (7) & (8)  \\ \hline 

Arp~2     &   0.999 &   0.027 &   0.920 &   0.011 &  13.415 &   0.826 &   0.45  \\
IC\,4499  &   0.976 &   0.016 &   0.902 &   0.007 &  13.460 &   0.620 &   0.30  \\
Lynga~7   &   1.321 &   0.042 &   1.158 &   0.016 &   7.829 &   0.587 &   0.40  \\
NGC\,104  &   1.397 &   0.017 &   1.162 &   0.006 &   6.860 &   0.540 &   0.44  \\
NGC\,362  &   1.069 &   0.009 &   0.953 &   0.005 &  11.578 &   0.568 &   0.52  \\
NGC\,1261 &   1.092 &   0.024 &   0.977 &   0.007 &  11.190 &   0.622 &   0.53  \\
NGC\,1851 &   1.110 &   0.007 &   0.999 &   0.005 &  11.059 &   0.545 &   0.53  \\
NGC\,2298 &   1.054 &   0.019 &   0.969 &   0.010 &  13.736 &   0.762 &   0.35  \\
NGC\,2808 &   1.077 &   0.008 &   0.967 &   0.004 &  11.215 &   0.461 &   0.56  \\
NGC\,3201 &   1.048 &   0.012 &   0.973 &   0.005 &  11.307 &   0.512 &   0.30  \\
NGC\,4590 &   0.950 &   0.012 &   0.896 &   0.006 &  13.559 &   0.505 &   0.30  \\
NGC\,4833 &   0.976 &   0.004 &   0.916 &   0.003 &  13.888 &   0.476 &   0.30  \\
NGC\,5024 &   0.975 &   0.002 &   0.897 &   0.002 &  12.324 &   0.525 &   0.20  \\
NGC\,5139 &   1.044 &   0.011 &   0.960 &   0.006 &  11.553 &   0.521 &   0.40  \\
NGC\,5272 &   1.018 &   0.003 &   0.929 &   0.002 &  12.281 &   0.528 &   0.40  \\
NGC\,5286 &   0.985 &   0.005 &   0.913 &   0.003 &  14.017 &   0.554 &   0.40  \\
NGC\,5904 &   1.057 &   0.005 &   0.962 &   0.003 &  11.818 &   0.523 &   0.50  \\
NGC\,5927 &   1.883 &   0.091 &   1.353 &   0.025 &   4.849 &   0.680 &   0.51  \\
NGC\,5986 &   1.019 &   0.006 &   0.939 &   0.004 &  12.747 &   0.553 &   0.40  \\
NGC\,6093 &   1.053 &   0.006 &   0.968 &   0.006 &  12.131 &   0.589 &   0.40  \\
NGC\,6101 &   1.024 &   0.018 &   0.956 &   0.007 &  13.104 &   0.698 &   0.30  \\
NGC\,6121 &   1.179 &   0.028 &   1.068 &   0.015 &  10.401 &   0.849 &   0.30  \\
NGC\,6144 &   1.028 &   0.014 &   0.961 &   0.009 &  14.346 &   0.962 &   0.25  \\
NGC\,6171 &   1.231 &   0.017 &   1.108 &   0.017 &   9.498 &   0.622 &   0.42  \\
NGC\,6205 &   1.038 &   0.008 &   0.948 &   0.003 &  11.467 &   0.509 &   0.40  \\
NGC\,6218 &   1.119 &   0.015 &   1.003 &   0.007 &   9.887 &   0.583 &   0.30  \\
NGC\,6254 &   1.061 &   0.021 &   0.972 &   0.009 &  11.391 &   0.524 &   0.50  \\
NGC\,6304 &   2.051 &   0.176 &   1.392 &   0.043 &   4.257 &   0.848 &   0.49  \\
NGC\,6341 &   0.956 &   0.003 &   0.889 &   0.003 &  13.381 &   0.491 &   0.30  \\
NGC\,6362 &   1.186 &   0.023 &   1.034 &   0.009 &   9.554 &   0.571 &   0.54  \\
NGC\,6426 &   1.008 &   0.027 &   0.946 &   0.026 &  13.152 &   1.967 &   0.40  \\
NGC\,6496 &   1.587 &   0.078 &   1.308 &   0.036 &   5.976 &   0.772 &   0.53  \\
NGC\,6541 &   0.954 &   0.004 &   0.883 &   0.003 &  12.937 &   0.523 &   0.30  \\
NGC\,6584 &   1.015 &   0.005 &   0.925 &   0.003 &  12.120 &   0.557 &   0.40  \\
NGC\,6624 &   1.514 &   0.096 &   1.215 &   0.040 &   6.412 &   0.742 &   0.47  \\
NGC\,6637 &   1.427 &   0.023 &   1.159 &   0.009 &   7.243 &   0.675 &   0.46  \\
NGC\,6652 &   1.321 &   0.021 &   1.147 &   0.014 &   7.829 &   0.564 &   0.47  \\
NGC\,6656 &   1.014 &   0.012 &   0.949 &   0.005 &  13.284 &   0.523 &   0.50  \\
NGC\,6681 &   1.087 &   0.010 &   0.983 &   0.005 &  12.476 &   0.579 &   0.60  \\
NGC\,6717 &   1.091 &   0.062 &   0.995 &   0.041 &  11.462 &   1.647 &   0.60  \\
NGC\,6723 &   1.149 &   0.022 &   1.033 &   0.010 &  10.487 &   0.574 &   0.49  \\
NGC\,6752 &   1.075 &   0.025 &   0.980 &   0.010 &  10.063 &   0.564 &   0.25  \\
NGC\,6779 &   0.960 &   0.005 &   0.888 &   0.002 &  13.579 &   0.519 &   0.40  \\
NGC\,6809 &   1.011 &   0.007 &   0.938 &   0.006 &  13.041 &   0.638 &   0.40  \\
NGC\,6934 &   1.050 &   0.018 &   0.962 &   0.015 &  10.874 &   1.048 &   0.40  \\
NGC\,6981 &   1.015 &   0.009 &   0.933 &   0.004 &  12.173 &   0.559 &   0.40  \\
NGC\,7006 &   1.035 &   0.014 &   0.955 &   0.006 &  12.705 &   0.575 &   0.40  \\
NGC\,7078 &   0.927 &   0.002 &   0.861 &   0.002 &  13.877 &   0.520 &   0.30  \\
NGC\,7089 &   0.996 &   0.005 &   0.909 &   0.006 &  12.324 &   0.615 &   0.30  \\
NGC\,7099 &   0.978 &   0.005 &   0.908 &   0.003 &  13.211 &   0.475 &   0.40  \\
Pal~2     &   1.019 &   0.015 &   0.962 &   0.010 &  11.582 &   0.558 &   0.30  \\
Rup~106   &   0.965 &   0.014 &   0.890 &   0.009 &  13.509 &   0.873 &   0.30  \\
Terzan~8  &   0.990 &   0.023 &   0.905 &   0.011 &  13.440 &   0.872 &   0.30  \\   
\end{tabular}
\tablefoot{Keys to columns: (1) name;  (2)-(3) RGB color at M$_I$=-3.5 and its uncertainty;  (4)-(5) RGB color at M$_I$=-3.0 and its uncertainty; (6)-(7) S-index and its uncertainty (8) absolute magnitude of the horizontal branch used for the determination of the S-index.
Note that we use V and I as shorthand for F606W and F814W respectively here.} 
\label{GC_results}
\end{table*}

\end{document}